\documentclass[fleqn,usenatbib]{mnras}

\usepackage{newtxtext,newtxmath}

\usepackage{graphicx}
\usepackage{dcolumn}
\usepackage{slashbox}

\usepackage[usenames,dvipsnames]{xcolor}
\hypersetup{colorlinks=true,citecolor=blue,linkcolor=OliveGreen}

\usepackage{soul}

\usepackage{subfigure}
\usepackage{wasysym}
\usepackage{verbatim}
\usepackage{color}
\usepackage{mathtools}
\usepackage{hyperref}
\usepackage{soul}
\usepackage{pbox}
\usepackage[export]{adjustbox}

\usepackage{slashed}
\usepackage{multirow}
\usepackage{booktabs}  
\usepackage{natbib}
\usepackage{float}
\usepackage{afterpage}
\usepackage{dcolumn}
\usepackage{bm}
\usepackage{lipsum}
\usepackage{setspace}
\usepackage{etoolbox}

\newcommand{\Abacus}[1]{\textsc{#1}}

\def\beq{\begin{equation}}
\def\eeq{\end{equation}}

\newlength{\depthofsumsign}
\newcommand{\nsum}[1][1.4]{
    \mathop{%
        \raisebox
            {-#1\depthofsumsign+1\depthofsumsign}
            {\scalebox
                {#1}
                {$\displaystyle\sum$}%
            }
    }
}

\def\mnras{Mon. Not. R. Astron. Soc}
\def\physrep{Physics Reports}

\usepackage[T1]{fontenc}
\usepackage{ae,aecompl}

\usepackage{graphicx}	
\usepackage{amsmath}	

\usepackage{lineno}

\title[The halo light cone catalogues of \Abacus{AbacusSummit}]{The halo light cone catalogues of \Abacus{AbacusSummit}}

\author[B. Hadzhiyska et al.]{
Boryana Hadzhiyska,$^{1}$\thanks{E-mail: boryana.hadzhiyska@cfa.harvard.edu}
Lehman H. Garrison$^{2}$,
Daniel Eisenstein$^{1}$,
and Sownak Bose$^{1,3}$
\\
$^{1}$Harvard-Smithsonian center for Astrophysics, 60 Garden St., Cambridge, MA 02138, USA\\
$^{2}$Center for Computational Astrophysics, Flatiron Institute, 162 Fifth Avenue, New York, NY 10010, USA\\
$^{3}$Institute for Computational Cosmology, Department of Physics, Durham University, Durham DH1 3LE, UK\\
}
\date{Accepted XXX. Received YYY; in original form ZZZ}

\pubyear{2021}

\begin{document}
\label{firstpage}
\pagerange{\pageref{firstpage}--\pageref{lastpage}}
\maketitle

\begin{abstract}
We describe a method for generating halo catalogues on the light cone using the \Abacus{AbacusSummit} suite of $N$-body simulations. The main application of these catalogues is the construction of realistic mock galaxy catalogues and weak lensing maps on the sky. Our algorithm associates the haloes from a set of coarsely-spaced snapshots with their positions at the time of light-cone crossing by matching halo particles to on-the-fly light cone particles. It then records the halo and particle information into an easily accessible product, which we call the \Abacus{AbacusSummit} halo light cone catalogues. Our recommended use of this product is in the halo mass regime of $M_{\rm halo} > 2.1 \times 10^{11} \ M_\odot/h$ for the \texttt{base} resolution simulations, i.e. haloes containing at least 100 particles, where the interpolated halo properties are most reliable. To test the validity of the obtained catalogues, we perform various visual inspections and consistency checks. In particular, we construct galaxy mock catalogues of emission-line galaxies (ELGs) at $z \sim 1$ by adopting a modified version of the \Abacus{AbacusHOD} script, which builds on the standard halo occupation distribution (HOD) method by including various extensions. We find that the multipoles of the auto-correlation function are consistent with the predictions from the full-box snapshot, implicitly validating our algorithm. In addition, we compute and output CMB convergence maps and find that the auto- and cross-power spectrum agrees with the theoretical prediction at the subpercent level.

Halo light cone catalogues for 25 \texttt{base} and 2 \texttt{huge} simulations at the fiducial cosmology is available at DOI:\href{https://www.doi.org/10.13139/OLCF/1825069}{10.13139/OLCF/1825069}
\end{abstract}




\section{Introduction}
\label{sec:intro}
In the near future, galaxy surveys such as the Dark Energy Spectroscopic Instrument (DESI) survey \citep{2016arXiv161100036D} and \textit{Euclid} \citep{2011arXiv1110.3193L} will measure the expansion history of the Universe and the growth of cosmic structure, which will enable us to understand the nature of dark matter and dark energy, and place constraints on cosmological parameters. This progress will be achieved through precise measurements of the galaxy clustering, baryon acoustic oscillations (BAO) peak, and weak lensing. However, in order to reach this level of precision, it is necessary to study and quantify the systematic uncertainties present in those surveys, which calls for the development of accurate mock catalogues \citep{2008RSPTA.366.4381B}. One benefit of using a mock catalogue is that the `true' value of any statistic can be measured directly and unambiguously, so we can check how reliable our tools for analyzing real observations are. Another benefit is that mocks allow us to test observational strategies and quantify levels of sample incompleteness. In many cases, it is not possible to assign a fibre to every galaxy due to mechanical constraints \cite[e.g.][]{2003MNRAS.346...78H,2012ApJ...756..127G,2017JCAP...03..001B,2017MNRAS.467.1940H,2017JCAP...04..008P} and even if a fibre is assigned, the redshift measurement can fail if the emission lines of the galaxy or the surface brightness are not strong enough. An incomplete sample may affect significantly the inferred clustering measurements, and in order to mitigate its effects on the measured signal, we need to model it reliably and in detail. 

Mock catalogues with realistic galaxy populations can be generated by using cosmological $N$-body simulations. In order to obtain a measurement of the BAO peak, which is crucial for upcoming galaxy surveys, the volumes of these simulations needs to be a few Gpc$^3$. While producing a hydrodynamical simulation of that volume at the required level of resolution is currently unfeasible due to the computational expense, dark-matter-only simulations are much less expensive. The downside of using such simulations is that one needs to adopt a scheme for ``painting'' galaxies on top of the dark-matter haloes. Several well-known population mechanisms are often adopted: the halo occupation distribution (HOD) \cite[e.g.][]{2000MNRAS.318.1144P,2000MNRAS.318..203S,2001ApJ...546...20S,2002ApJ...575..587B,2004ApJ...609...35K,2005ApJ...633..791Z}, which describes the probability a halo with mass $M_{\rm halo}$ contains $N_g$ galaxies; subhalo abundance matching (SHAM) \cite[e.g.][]{2004MNRAS.353..189V,2007ApJ...668..826C}, which relates directly subhalo properties (such as mass and circular velocity) to galaxy properties (such as luminosity and stellar mass); and semi-analytic models (SAMs) \cite[e.g.][]{2006RPPh...69.3101B,2012NewA...17..175B,2008MNRAS.391..481S}, which uses analytic prescriptions to model the formation and evolution of galaxies by usually utilizing the halo merger histories.

A prerequisite for applying a SAM to a simulation is the existence of high-resolution merger trees, which are difficult to construct for large simulations. Nevertheless, there are approaches which can augment the resolution of the simulation merger trees \cite[e.g.][]{2013MNRAS.435..743D,2014MNRAS.442.3256A,2016ComAC...3....3B}, but a lot of questions regarding the tuning of the various SAM parameters are still poorly understood. The SHAM prescription, on the other hand, requires a complete subhalo catalogue, which can be hard to obtain, since smaller-mass subhaloes often undergo multiple mergers, and the ability of a subhalo to survive a merger is strongly dependent on the resolution of the simulation. Finally, the HOD prescription can be applied to lower-resolution simulations because satellite galaxies do not need to be placed on subhaloes but instead, can either follow an analytical distribution or the dark-matter particle distribution \cite[e.g.][]{2010MNRAS.405..143A}.

The ideal procedure for emulating observations involves populating haloes with galaxies on a light cone that has been directly output from a simulation. However, most simulations only output snapshots at discrete times and thus, apply the HOD method to a single snapshot. The downside of using discrete time epochs is that the halo bias is constant in time rather than evolving with redshift, which affects the clustering measurements. To ameliorate that effect, one could join together multiple snapshots, but there would be discontinuities at the boundaries, and objects may appear multiple times at the boundary or be completely missing \cite[e.g.][]{2015MNRAS.447.1319F}. Moreover, the baseline HOD and SHAM techniques do not take into consideration evolution of the galaxy population, although there have been attempts in SHAM modeling that aim to, e.g., reproduce the observed stellar mass function at different redshifts \citep{2010ApJ...710..903M}. As for the HOD method, there is currently no robust way to model its evolution, as it is strongly dependent on the type of galaxies under study. For example, \cite{2020JCAP...03..044N} allow for time evolution of the HOD parameters by Taylor expanding them as a function of scale factor and constraining both their fiducial values as well as their first derivatives. \citet{Karim+2021} adopt an interpolation scheme for obtaining the HOD parameters at different redshifts, while \cite{2015MNRAS.452.1861C} use SAMs to parametrize their evolution.

In this paper, we describe the halo light cone catalogues of the \Abacus{AbacusSummit} simulation, which are designed for generating realistic mock catalogues that meet the requirements of current galaxy surveys. Our procedure involves the following steps. First, we construct a halo light cone catalogue from the simulation by interpolating the positions of haloes between snapshots, using merger tree information. We then find the subsampled particles belonging to the haloes in the on-the-fly particle light cones, obtaining a light cone catalogue with haloes and particles. We recommend using this product in the halo mass regime of $M_{\rm halo} > 2.1 \times 10^{11} \ M_\odot/h$, corresponding to haloes with 100 or more particles. This catalogue can then be easily populated with galaxies using an HOD prescription, reproducing the observed effect of measuring clustering on the sky. In an accompanying paper \citep{Yuan+2021}, an augmented HOD method for populating dark-matter haloes from the \Abacus{AbacusSummit} simulation is described. 

This paper is organized as follows: Section \ref{sec:abacus} presents the \Abacus{AbacusSummit} simulation and accompanying products, such as particle subsamples, merger trees, and light cones. Section \ref{sec:light} outlines the method for generating the halo light cone catalogues, and Section \ref{ssec:tests} tests it. We describe some potential applications of the halo light cone catalogues in Section \ref{ssec:mock} by comparing mock catalogues obtained on the light cone with those coming from a full simulation snapshot and testing predictions of cross-correlation measurements with CMB lensing. Section \ref{sec:conc} summarizes our main results and conclusions.

\section{The \Abacus{AbacusSummit} simulation}
\label{sec:abacus}
The \Abacus{AbacusSummit} suite of high-performance cosmological $N$-body simulations \citep{Maksimova+2021} is designed to meet the Cosmological Simulation Requirements of the Dark Energy Spectroscopic Instrument (DESI) survey and run on the Summit supercomputer at the Oak Ridge Leadership Computing Facility. The simulations are run with \Abacus{Abacus} \citep{2019MNRAS.485.3370G,Garrison+2021b}, a high-accuracy cosmological $N$-body simulation code, which is optimized for GPU architectures and  for large-volume, moderately clustered simulations. \Abacus{Abacus} is extremely fast, performing 70 million particle updates per second on each node of the Summit supercomputer, and also extremely accurate, with typical force accuracy below $10^{-5}$. The near-field computations run on a GPU architecture, whereas the far-field computations run on CPUs. {For full details on all data products, see \citet{Maksimova+2021}.}

{The \Abacus{AbacusSummit} halo light cone catalogues are available at DOI:XXXXX for the 25 \texttt{base} simulations (\texttt{AbacusSummit\_base\_c000\_ph\{000-024\}}) and the two \texttt{huge} simulations (\texttt{AbacusSummit\_huge\_c000\_ph\{201,202\}}) at the fiducial \Abacus{AbacusSummit} cosmology: $\Omega_b h^2 = 0.02237$, $\Omega_c h^2 = 0.02237$, $h = 0.6736$, $10^9 A_s = 2.0830$, $n_s = 0.9649$, $w_0 = -1$, $w_a = 0$. The box sizes of the \texttt{base} and \texttt{huge} simulations are $2000 \ {\rm Mpc}/h$ and $7600 \ {\rm Mpc}/h$, respectively, whereas the particle masses are $M_{\rm part} = 2.1 \times 10^{9} \ M_\odot/h$ and $M_{\rm part} = 2.1 \times 5 \times 10^{10} \ M_\odot/h$ corresponding to 6912$^3$ and 8640$^3$ particles. The halo light cone catalogues have been designed with the aim of supporting the construction of mock catalogues using halo occupation distributions and enabling efficient access to measurements of the density fields. The \texttt{base} catalogues are generated for the redshift epochs: $z = 0.1, \ 0.15, \ 0.2, \ 0.25, \ 0.3, \ 0.35, \ 0.4, \ 0.45, \ 0.5, \ 0.575, \ 0.65, \ 0.725, \\ \ 0.8, \ 0.875, \ 0.95, \ 1.025, \ 1.1, \ 1.175, \ 1.25, \ 1.325, \ 1.4, \ 1.475, \ 1.55, \\ \ 1.625, \ 1.7, \ 1.775, \ 1.85, \ 1.925, \ 2.0, \ 2.25, \ 2.5$. And the \texttt{huge} catalogues are available for all epochs until $z = 2.25$. In the next sections, we summarize the \Abacus{AbacusSummit} products that we utilized in the creation of the halo light cone catalogues.}

\subsection{Particle subsamples}
\label{ssec:subsamp}

Particle subsamples are available at 12 primary and 21 secondary redshift epochs. They are composed of two subsamples, \texttt{A} and \texttt{B}, covering 3\% and 7\% of all particles, respectively. The subsampled particles are selected randomly and are consistent across redshift, which enables us to associate halo catalogues to the light cones and also to build merger trees. The intended use of the subsamples is to serve as proxies for satellite galaxies in galaxy-halo models. 

The 12 designated primary epochs are $z = 0.1, \ 0.2, \ 0.3, \ 0.4, \ 0.5, \ 0.8, \ 1.1, \ 1.4, \ 1.7, \ 2.0, \ 2.5, \ {\rm and} \ 3.0$. At these epochs, we output the positions, velocities and ID's of the subsamples. For the 21 secondary redshifts, covering redshifts $z = 0.15 - 8.0$, we only output the particle ID's. The particle ID's also encode information about the Lagrangian positions of the particles, their ``local density'' and whether they have been part of the largest L2 halo (within an L1).

\subsection{Light cones}
\label{ssec:cones}

\begin{figure}
    \centering
    \includegraphics[width=0.48\textwidth]{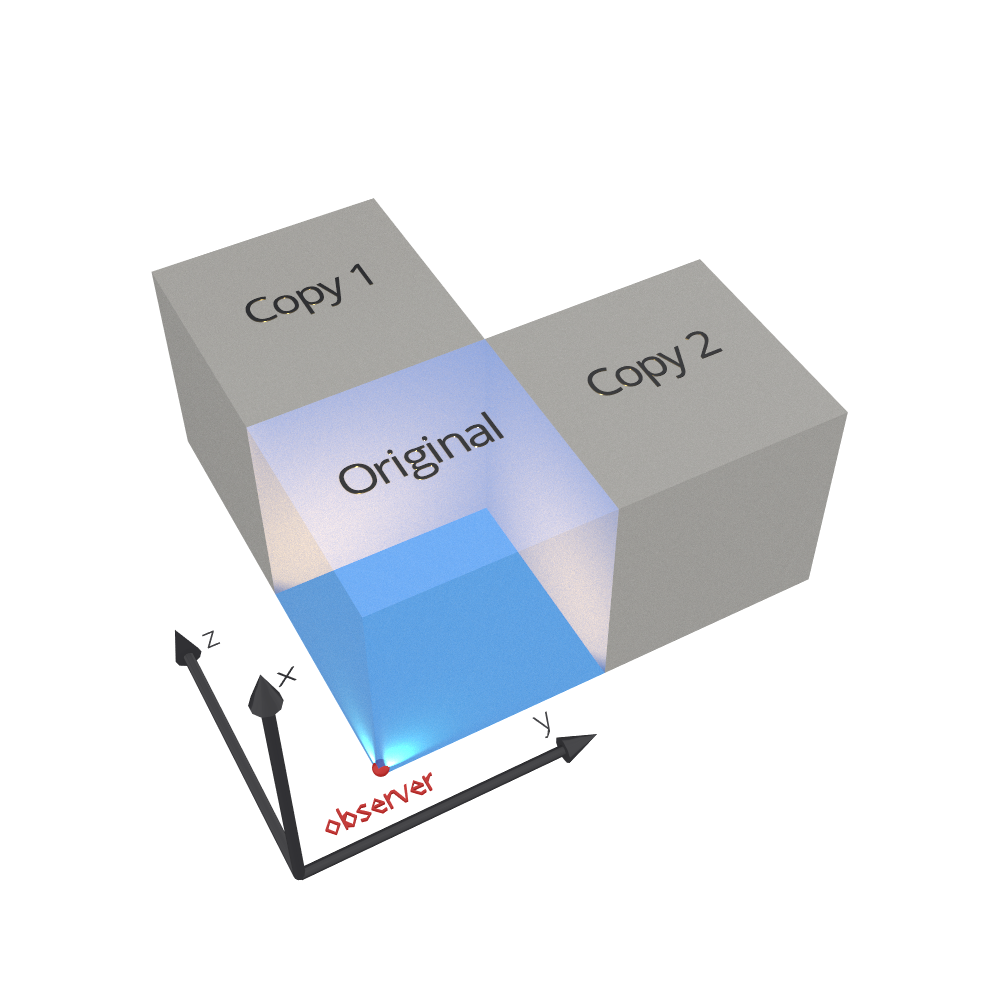} 
    \caption{Visualization of the geometrical arrangement of the \Abacus{AbacusSummit} light cones. The original box, of length 2000 Mpc$/h$, is centered at (0, 0, 0), while two identical copies are placed at (0, 0, 2000) and (0, 2000, 0) Mpc$/h$. The observer is located at the corner of the original box, at (-990, -990, -990) Mpc$/h$.} 
    \label{fig:geo}
\end{figure}

The \texttt{base} resolution \Abacus{AbacusSummit} boxes include a light cone stretching from the corner of the main box and two periodic copies of that box, seamlessly attached to it in the $y$ and $z$ directions (see Fig. \ref{fig:geo}). The exception to that are the \Abacus{AbacusSummit} \texttt{huge} boxes, which place the observer at the centre of the box and utilize only a single copy of the box. At the \texttt{base} resolution of the \Abacus{AbacusSummit} boxes (6912$^3$ particles, 2 Gpc$/h$ box), the halo light cone catalogues cover an octant of the sky to $z \approx 0.8$ and about 1800 deg$^2$ extending to $z \approx 2.45$. The \texttt{huge} boxes (8640$^3$ particles, 7.5 Gpc$/h$ box) provide light cone information of the full sky until $z \approx 2.18$ and extend further towards the corners of the box. 

At every timestep, \Abacus{Abacus} identifies particles that belong to the light cone and outputs their positions, velocities, particle IDs, and HEALPix pixel number, which can be used to form projected density maps. The pixel orientation is such that the $+z$ direction coincides with the North Pole. The HEALPix maps are output from all particles with resolution of $N_{\rm side} = 16384$, which is more than sufficient for performing accurate weak lensing analysis, whereas the particle outputs contain only a 10\% subsample of the particles, the so-called \texttt{A} and \texttt{B} subsamples (see Section \ref{ssec:subsamp}). Storing all particles in the form of high-resolution HEALPix maps required setting aside only 0.3 bytes per particle after compression, which is a much smaller expense than storing all particles individually. For more details on these products, we refer the reader to \citet{Maksimova+2021}.

The geometrical arrangement of the light cones is shown in Fig. \ref{fig:geo}. For a simulation with box length of 2000 Mpc$/h$ on a side, the light cone observer is positioned at (-990, -990, -990), or, in other words, 10 Mpc$/h$ inside the corner of the original box. Three boxes form the eligible space of the light cone, centered at (0, 0, 0), (0, 0, 2000), and (0, 2000, 0), respectively (measured in Mpc$/h$ units). Particles are output from every time step, where their trajectories are linearly interpolated to find the time when the light cone intersects their paths. Their positions and velocities are updated to this time. This provides an octant to a distance of 1990 Mpc$/h$ ($z \approx 0.8$), shrinking to two patches each about 900 square degrees at a distance of 3990 Mpc$/h$ ($z \approx 2.45$). For the \texttt{huge} boxes ($N = 8640^3$ and $L_{\rm box} = 7500$ Mpc$/h$), the light cone is simply one copy of the box, centered at (0, 0, 0), providing a full-sky light cone to the the half-distance of the box (3.75 Gpc/$h$), and further toward the eight corners.

\begin{figure*}
    \centering
    \includegraphics[width=0.8\textwidth]{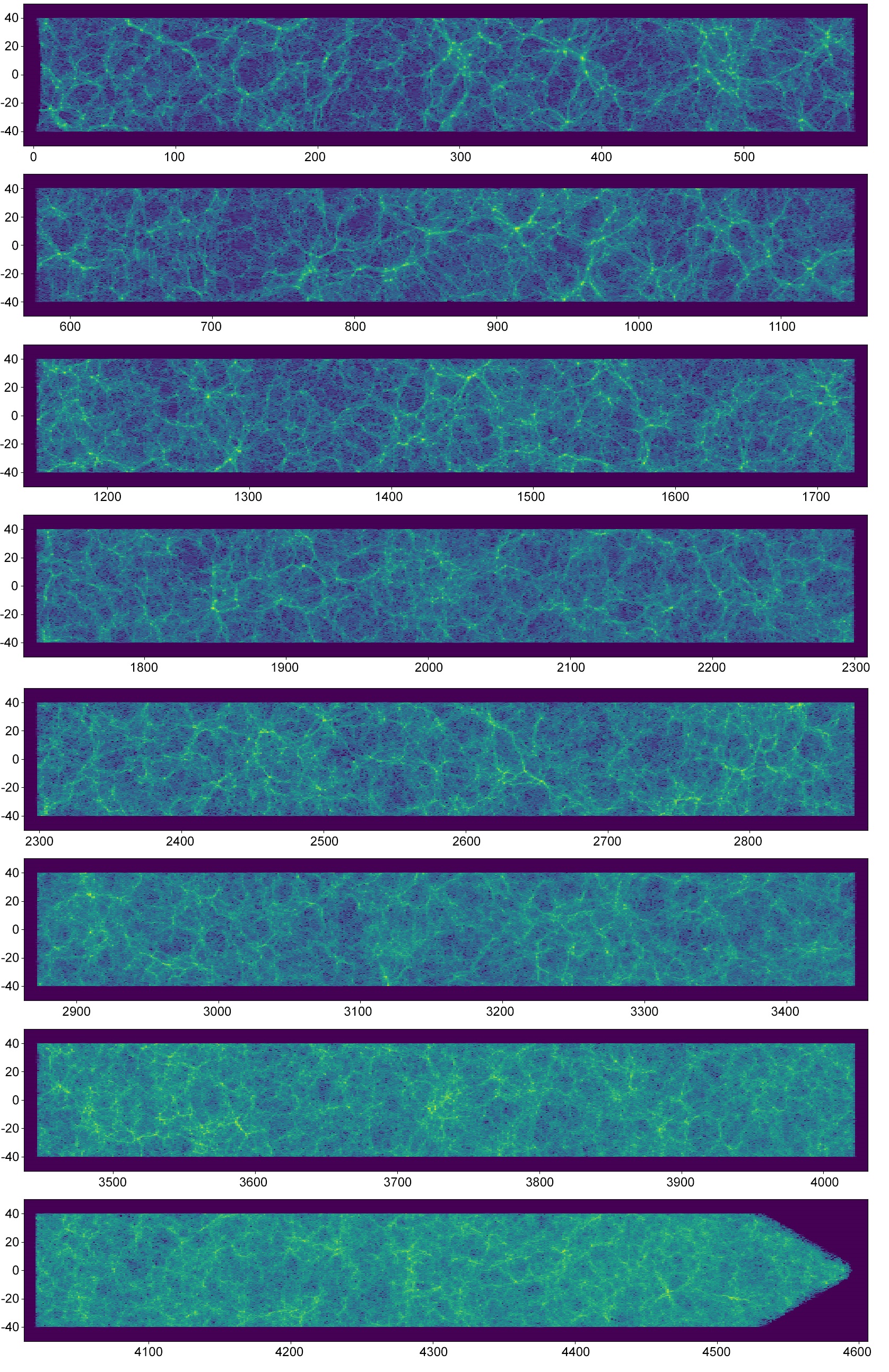}
    \caption{Visualization of the particle light cone, showing a narrow strip of width 80 Mpc$/h$ and thickness 10 Mpc$/h$, stretching from the corner of the original box, (-990, -990, -990) Mpc$/h$, to the corner of one of the box copies (1000, 3000, 1000) Mpc$/h$, and covering a comoving distance of nearly 4600 Mpc$/h$.} 
    \label{fig:light}
\end{figure*}

In Fig. \ref{fig:light}, we show a narrow strip of the light cone stretching from the corner of the original box (starting at $z = 0.1$) to the corner furthest from it, belonging to a copy of the box, at a distance of $\sim$ 4900 Mpc$/h$. The observer sits at the top left of the figure, where the structures seen have formed most recently. The bottom right corresponds to the most distant time epochs that have available light cone outputs.

\subsection{The \Abacus{CompaSO} halo finder}
\label{ssec:compaso}
Haloes in \Abacus{AbacusSummit} are defined using the  {\it Competitive Assignment to Spherical Overdensities} (\Abacus{CompaSO}) halo finder \citep{Hadzhiyska+2021}, an optimized on-the-fly method for identifying groups of particles in cosmological $N$-body simulations. \Abacus{CompaSO} builds upon existing spherical overdensity (SO) algorithms by taking into consideration the tidal radius around a smaller halo before competitively assigning halo membership to the particles. In this way, the \Abacus{CompaSO} finder allows for more effective deblending of haloes in close proximity of each other as well as the formation of new haloes on the outskirts of larger ones. It further requires that a particle that becomes a halo centre has the highest local density among its immediate  neighbors. \Abacus{CompaSO} is developed as a highly efficient on-the-fly group finder, which is crucial for enabling good load-balancing between the GPU and CPU and the creation of high-resolution merger trees. Halo finding is performed with a density threshold of 200 times the mean density, level-1 (L1), and 800 times the mean density, level-2 (L2). However, the \Abacus{AbacusSummit} halo catalogues store only L1 halo information. The centre-of-mass of the largest L2 subhalo is used to define the centre relative to which all L1 halo statistics are output. For a detailed description of the algorithm, performance tests and comparisons with other halo finders, see \citet{Hadzhiyska+2021}.

\subsection{Merger Trees}
\label{ssec:merger}

The merger tree method used in \Abacus{AbacusSummit} tracks the cores of haloes to determine associations between objects across multiple timeslices and records information about their progenitors and descendants. For details on the merger tree algorithm, we refer the reader to \citet{Bose+2021}.

For each halo, the method identifies the list of progenitor ({\tt Progenitor}) haloes as well as the main progenitor ({\tt MainProgenitor}) halo at the preceding redshift catalogue that share a substantial fraction of their particles with the halo at the present catalogue. We use the progenitor history of the halo to determine its trajectory through time and interpolate to get its position and velocity at the time of intersection with the observer's light cone.

\subsection{Cleaned halo catalogues}
\label{ssec:clean}

The choice of whether an object is identified as a halo by any halo-finding algorithm can be somewhat arbitrary. In the case of SO-based methods (as is the case for \Abacus{CompaSO}), the halo boundary is starkly determined by the SO threshold density, while for FoF-based finders, it is strongly dependent on the linking length parameter. This choice becomes even more challenging in dense regions of the simulation as well as for merging haloes and splashback events, where dynamical processes such as fly-bys, partial mergers, splits often take place.

To overcome these issues in \Abacus{AbacusSummit}, we have `removed' objects in the \Abacus{CompaSO} halo catalogues that may have been compromised. Haloes flagged as unphysical have their masses added to a massive neighbour that they share history with at the time they attain peak mass. The aggregated halo remains merged for all subsequent outputs, and the flagged halo is removed from the halo catalogue at each of the subsequent outputs. The particle list of the aggregate halo is the union of the particles of the original halo and all haloes that have been merged with it.

\subsection{The \Abacus{AbacusHOD} model}
\label{ssec:hod}

The \Abacus{AbacusSummit} halo light cone catalogues are designed to generate mock catalogues via the \Abacus{AbacusHOD} model, a sophisticated routine hat builds upon the baseline HOD model by incorporating various generalizations pertaining to halo-scale physics and assembly bias. \Abacus{AbacusHOD} allows the user to specify different tracer types: emission-line galaxies (ELGs), luminous red galaxies (LRGs), and quasistellar objects (QSOs). The \Abacus{AbacusHOD} model is based on the GeneRalized ANd Differentiable Halo Occupation Distribution (GRAND-HOD) model \cite[see][]{2018MNRAS.478.2019Y} and is described in detail in \citet{Yuan+2021}. 

The decorations parameters incorporated in the \Abacus{AbacusHOD} model are listed below:
\begin{itemize}
    \item \texttt{s}  is the satellite profile modulation parameter, which modulates how the radial distribution of satellite galaxies within haloes deviate from the radial profile of the halo.
    \item \texttt{s\_v} is the satellite velocity bias parameter, which modulates how the satellite galaxy peculiar velocity deviates from that of the local dark matter particle.
    \item \texttt{alpha\_c} is the central velocity bias parameter, which modulates the peculiar velocity of the central galaxy.
    \item \texttt{s\_p} is the perihelion distance modulation parameter.
    \item \texttt{A\_c} or \texttt{A\_s} are the concentration assembly bias parameters for centrals and satellites, respectively.
    \item \texttt{B\_c} or \texttt{B\_s} are the environment assembly bias parameters for centrals and satellites, respectively. To define halo environment, we adopt the same formalism as \citet{2020MNRAS.493.5506H}.
\end{itemize}

We note that the assembly bias implementation preserves the overall galaxy number density by reranking haloes based on their pseudo-mass.

\section{Algorithm}
\label{sec:light}

In this section, we describe the algorithm for obtaining the halo light cone catalogues of \Abacus{AbacusSummit} by combining information from the merger trees, cleaned \Abacus{CompaSO} halo catalogues, and the particle light cones. The final product includes various statistical properties of the haloes and spatial information about the particles, which can be used to generate mock catalogues. Here we describe the method for the \Abacus{AbacusSummit} \texttt{base} simulations, but the general approach remains the same for all boxes. 

\subsection{Theoretical pretext}
\label{ssec:pretext}

The intended application of the \Abacus{AbacusSummit} halo light cone catalogues is the construction of high-fidelity mocks for galaxy clustering and weak lensing surveys. For this reason, we take special care to output as accurately as possible the interpolated velocities and positions of the haloes, which are utilised as proxies for the central galaxies. The satellites are selected from the particle subsamples. An accurate placing of the centrals and satellites is key to modeling large-scale structure probes such as galaxy clustering and lensing measurements in redshift surveys. The catalogues are designed to support galaxy embeddings via the \Abacus{AbacusHOD} script (see Section~\ref{ssec:hod}), but can also be used independently.

One route for obtaining such catalogues is to run a halo finder on all particles on the light cone with a halo boundary that evolves with comoving distance. This requires storing a huge amount of particle data, which is infeasible for a simulation suite as large as \Abacus{AbacusSummit}. Another approach is to use the halo information from the full-box redshift catalogues to construct halo light cones. The simplest method to select haloes for the light cone is with a ``cookie cutter'', where the halo allegiance to a light cone catalogue is determined solely by its momentary position at some redshift epoch. However, this method suffers from several issues: In practice, the haloes are not stationary but instead traverse a non-negligible distance between any two redshift catalogues. Therefore, by assuming that they are at rest, we risk duplicating and missing haloes that reside on the boundary between two redshift epochs. In addition, this would also lead to an unphysically large difference between the halo position and the positions of its particles, which would in turn alter the one-halo term and the cross-correlation measurement between galaxies and the matter field.

The approach we adopt in \Abacus{AbacusSummit}, however, is more sophisticated, since the main application is the forward modeling of cosmological surveys. The aim of the algorithm is to find the position and properties of the haloes at the time of light cone crossing, using a combination of interpolation and particle matching. The interpolation is done using merger tree information: knowing a halo's position, velocity, and mass at the current epoch and its main progenitor in the previous epoch, one can interpolate these properties to the moment of light-cone crossing. While the particle light cones are output with excellent time granularity (every \textsc{Abacus} time step, i.e. $\Delta \log a \sim 0.001$), the redshift catalogues, and thus the merger tree associations, are output only at epochs corresponding to the primary and secondary redshifts. However, this output resolution is sufficient for reliably tracking the dense cores of haloes across time and finding their light-cone crossing times.

Before proceeding with the details of the algorithm, we offer an estimate of the error introduced by interpolation (we measure such effects in Section \ref{ssec:tests}).  The initial step is to linearly interpolate the halo positions and velocities between the coarser redshift catalogues, assuming that the halo trajectories are kinematic. The bulk velocity of a typical halo is highest, 500 km/s (0.5 kpc/Myr), at low redshifts, $z \lesssim 0.5$. The redshift catalogues there are spaced by $\Delta z = 0.05$ ($\Delta \log a \approx 0.04$, or $\sim$500 Myr), so on average the haloes would move by $\sim$300 kpc. The distance traveled is, thus, small enough that the assumption of linear motion does not contribute substantial bias to the interpolation. This is in fact a conservative estimate, as a fraction of the halo velocity comes from the infall of haloes into large clusters.

Alternatively, one can compare the halo crossing time to the time intervals between the redshift catalogues. The time interval between two redshift catalogues expressed in units of the redshift-dependent Hubble time is given by $\Delta t = \Delta \log{a}/H(a)$. The halo crossing time is given by $t_{\rm cross} = 2 R_{\rm vir}/V_{\rm vir}$, with $V_{\rm vir} = \sqrt{2 G M_{\rm vir}/R_{\rm vir}}$ and $M_{\rm vir} = 4/3 \pi R_{\rm vir}^3 \Delta_{\rm vir}(a) \rho_{\rm crit}(a)$, where $\rho_{\rm crit}(a) = 3 H(a)^2/(8 \pi G)$ is the redshift-dependent critical density of the Universe and $\Delta_{\rm vir} (a)$ is the virial density contrast, which in \Abacus{Abacus} is given by the fitting function \citep{1998ApJ...495...80B}: $\Delta_{\rm vir} (a) \equiv \Delta_{\rm L1} = (200/ 18 \pi^2) \ (18 \pi^2 + 82 x - 39 x^2)$ with $x = \Omega_m(a) - 1$ \citep{Hadzhiyska+2021}. Simplifying, we obtain $t_{\rm cross} = 2/(\sqrt{\Delta_{\rm vir}(a)} H(a))$. For $z \lesssim 0.5$, $\Delta \log{a} \approx 0.04$ and $\Delta_{\rm vir} (a) \approx 110$, so the time interval is $\Delta t \approx 0.04 / H(a)$ and the crossing time is $t_{\rm cross} \approx 0.2/H(a)$. Thus, the time interval between the catalogues is about 20\% of the crossing time of haloes, demonstrating that we retain accuracy in interpolating between the redshift epochs to track the motion of haloes.

\subsection{Selection of haloes}
\label{sssec:build}
The first step in this process is finding the haloes that intersect the observer's light cone for each redshift catalogue. This task is realized by using the halo catalogues and the halo merger trees, described in Section \ref{ssec:merger}. 

The halo light cone catalogues are segmented into redshift catalogues, starting at $z_1 = 0.1$ and finishing at $z_{33} = 8$. A catalogue at redshift $z_i$ corresponds to a comoving distance to the light cone $\chi_i$, with $\chi_i = \int_{t_i}^{t_0} c dt'/a(t')$, $t_0$ the present time, and $t_i \equiv t(z_i)$. We aim to select haloes that cross the light cone between $z_{i-1/2}$ and $z_{i+1/2}$. The ``half redshifts'', $z_{i\pm1/2}$, are defined as $z((\chi_i + \chi_{i\pm1})/2)$. We differentiate between haloes with merger tree information and haloes without merger tree information, i.e. haloes at $z_i$ whose main progenitor has been successfully identified in the previous catalogue, $z_{i+1}$.

When merger tree information is available, we compute the time at which each halo is crossed by the light cone to determine whether it belongs to the halo catalogue, $z_i$.
\begin{enumerate}
    \item We first compute the distance between each halo in the current epoch ($\chi_{i}$) and the light cone origin, $r_i$, the distance between the halo's main progenitor in the preceding epoch ($\chi_{i+1}$) and the origin, $r_{i+1}$, and the distance between the halo's descendant in the next epoch ($\chi_{i-1}$) and the origin, $r_{i-1}$.
    \item We determine the {comoving distances to the light cone crossing time}, $\chi_{\ast}$, for all haloes in the catalogue by assuming that they move radially with a constant velocity given by $(r_{i+1}-r_{i})/(\chi_{i}-\chi_{i+1})$. 
    
    For haloes with light cone crossing times between $\chi_{i} < \chi_\ast < \chi_{i+1/2}$, the interpolation is obtained via:
    \begin{equation}
        r_{i+1}+\left(\frac{\chi_{\ast}-\chi_{i+1}}{\chi_{i}-\chi_{i+1}}\right) (r_{i+1}-r_{i}) = \chi_{\ast},
    \end{equation}
    whereas if $\chi_{i-1/2} \leq \chi_\ast \leq \chi_{i}$, it is
    \begin{equation}
        r_{i}+\left(\frac{\chi_{\ast}-\chi_{i}}{\chi_{i-1}-\chi_{i}}\right) (r_{i}-r_{i-1}) = \chi_{\ast}.
    \end{equation}
    \item Once we have determined the comoving distance $\chi_\ast$, we interpolate to find the mass, position and velocity of the halo at the time of crossing, assuming that the halo moves in a straight line at a constant velocity between the two epochs and that its mass changes linearly:
    \begin{eqnarray}
        q_{\rm interp} = q_i + \left(\frac{\chi_{\ast}-\chi_{i+1}}{\chi_{i}-\chi_{i+1}}\right) (q_{i+1}-q_i), \ \ \ {\rm if} \ \chi_{i} < \chi_\ast < \chi_{i+1/2} \ \nonumber \\
        q_{\rm interp} = q_i + \left(\frac{\chi_{\ast}-\chi_{i}}{\chi_{i-1}-\chi_{i}}\right) (q_{i}-q_{i-1}), \ \ \ {\rm if} \ \chi_{i-1/2} \leq \chi_\ast \leq \chi_{i},
    \end{eqnarray}
    where $q = \{ \mathbf{x}, \mathbf{v}, M \}$ can stand for position, velocity or mass.
    \item In order to avoid the duplication of particles in multiple haloes, before proceeding with the next redshift, $z_{i+1}$, we mark ineligible for future consideration the progenitors of all selected haloes in $z_i$.
\end{enumerate}

\begin{figure}
    \centering
    \includegraphics[width=0.48\textwidth]{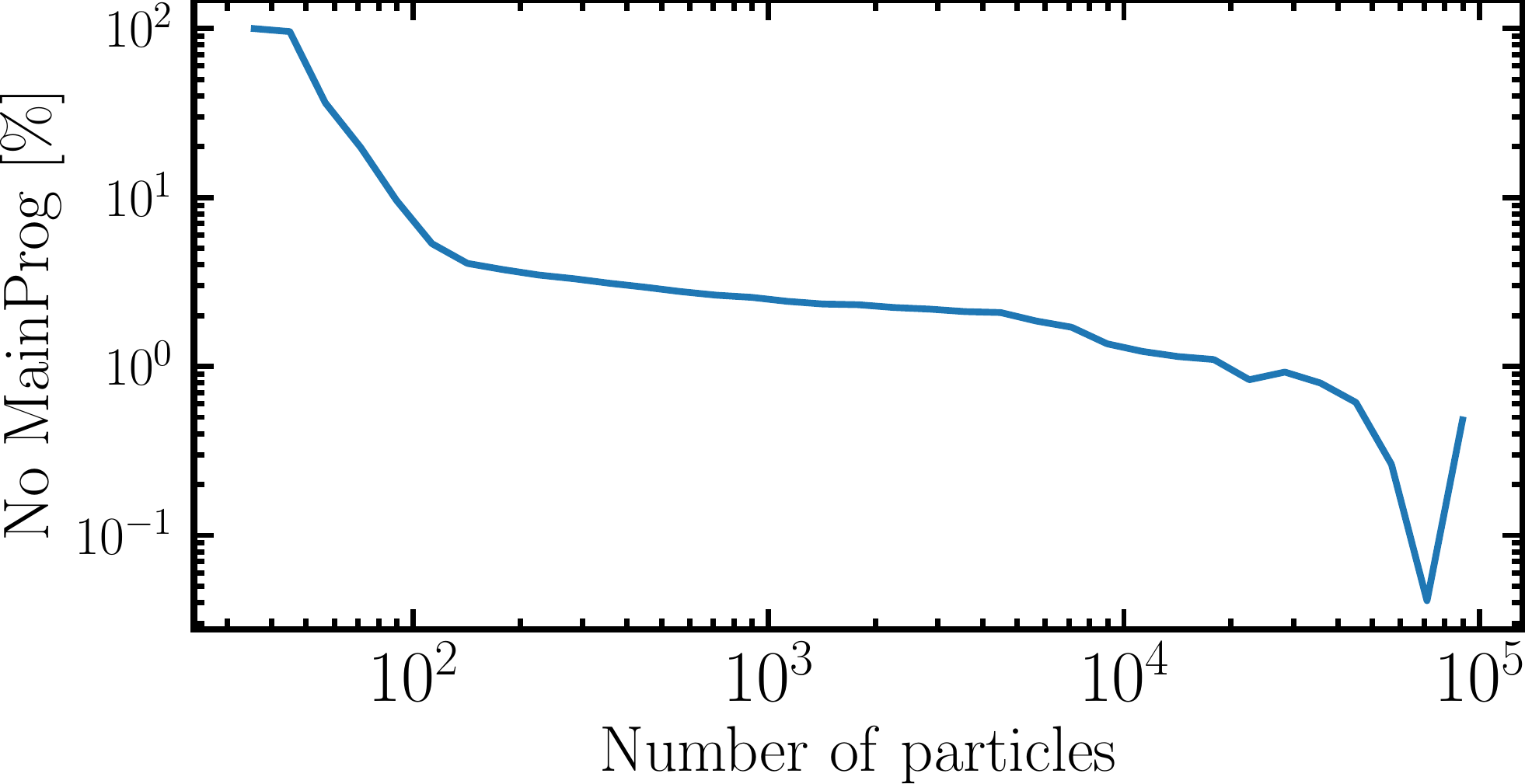}
    \caption{Percentage of haloes at $z = 0.8$ whose main progenitor in the preceding redshift ($z = 0.875$) is not identified. The effect is most pronounced for small-mass haloes and about 5\% of the haloes with $\sim$100 particles, 2.5\% of the haloes with $\sim$1000 particles, and 1\% of the haloes with $\sim$10000 particles.}
    \label{fig:no_id}
\end{figure}

When no merger tree information is available, we only select haloes whose distances to the observer at the current redshift, $z_{i}$, lie between $z_{i-1/2}$ and $z_{i+1/2}$. Their interpolated velocities and masses are copied over from their ``stationary'' equivalents at $z_i$, whereas the interpolated (comoving) position is obtained via:
 \begin{equation}
     \mathbf{x}_{\rm interp} = \mathbf{x}_i + \mathbf{v}_i \ [t(r_i)-t(\chi_i)] \ (1 + z_i), 
 \end{equation}
 where $t(r_i)$ is the proper time corresponding to the comoving distance of each halo to the light cone origin and $t(\chi_i)$ is the proper time of the $i$-th redshift catalogue. At a given redshift, the objects without merger associations make up about 1/3 of all haloes (since a large portion of the haloes are small and hard to track in time), but only about 5\% of the haloes with $\sim$100 particles, 2.5\% of the haloes with $\sim$1000 particles, and 1\% of the haloes with $\sim$10000 particles (see Fig.~\ref{fig:no_id}).
 
To clean up the selection of haloes, we follow these additional steps:
\begin{itemize}
    \item We discard haloes and their particles from the edges of the light cones, since they cannot reliably be matched to the light cone particles due to the periodic boundary conditions. The light cone edges are defined as $10 \ {\rm Mpc}/h$ within the boundaries of the light cone boxes, i.e. we retain haloes with coordinates $-990 \ {\rm Mpc}/h < x < 990 \ {\rm Mpc}/h$, $ -990 \ {\rm Mpc}/h < y < 3000 \ {\rm Mpc}/h$, and $ -990 \ {\rm Mpc}/h < z < 3000 \ {\rm Mpc}/h$.
    \item We also remove all repeated haloes and their particles (about 0.4\% overall), keeping only the first instance of each halo. These instances occur because at any redshift catalogue there is a small number of instances where haloes point to the same main progenitor. This is typically the case for the most massive haloes (about 25\% of the haloes with 10000 particles), since they have more complex substructure that occasionally gets arbitrarily segmented into several distincts haloes. Special care is taken for haloes on the boundary between two boxes (i.e. the original one and each of the two copies), where we keep the unique haloes located closer to the observer at (-990, -990, -990) ${\rm Mpc}/h$. Note that this choice preserves the small fraction of overlapping haloes in the two copies of the original box, which inevitably appear in redshift catalogues beyond $z \gtrsim 0.8$ due to the geometry of the light cones.
    \item Finally, we clean the haloes that have been deemed ``unphysical'' by the merger tree cleaning algorithm (see Section \ref{ssec:clean}) and record the halo properties of the surviving $\Abacus{CompaSO}$ objects. We thus eliminate all the `removed' haloes and add their particles to a more massive companion that they share recent merger history with. 
\end{itemize}

\subsection{Matching particle lists to haloes}
\label{sssec:match}
Associating particle lists with the halo catalogues is useful for providing a proxy for the placement of satellite galaxies in a mock catalogue and also for refining and testing our interpolation technique by examining the spatial distribution of the particles. We match the subsample \texttt{A} and \texttt{B} particles (3\% and 7\%, respectively) belonging to the selected haloes from Section \ref{sssec:build} to the light cone outputs. 
Note that the light cone outputs are available for all three boxes (see Fig.~\ref{fig:geo}), but haloes in catalogues with $z < 0.8$ come solely from the original box. We record the position and velocity on the light cone of each matched particle in the halo catalogues. We do not include in the halo light cone catalogues the particles that have no matches in the light cone catalogues. The unmatched particles constitute $\sim$0.0003\% of the total, and upon visualizing them, we find that they are located exactly at the boundary between two light cone shells. The most likely explanation for why those particles are missing is that they have traversed a larger than anticipated distance in the radial direction (relative to the observer), jumping to the next shell and escaping the output condition of the \Abacus{Abacus} code. Since those events are extremely rare and affect all haloes equally (regardless of their mass, position on the sky, etc.), we do not expect them to cause any substantial problems with the generated mock catalogues and the inferred large-scale structure probes.

\subsection{Data format}
\label{sssec:save}
The data for each halo light cone catalogue is stored in two files, \texttt{lc\_halo\_info.asdf} and  \texttt{lc\_pid\_rv.asdf}:
\begin{itemize}
    \item \texttt{lc\_halo\_info.asdf} contains the summary statistics of the haloes. In addition to the standard halo properties computed for the largest L2 subhalo (see Section \ref{ssec:compaso}), which are documented in \citet{Maksimova+2021}, we record various interpolated quantities: the number of particles, \texttt{N\_interp}; the centre of mass position of the largest L2 subhalo, \texttt{pos\_interp}; the centre of mass velocity of the largest L2 subhalo, \texttt{vel\_interp}; the averaged positions and velocities of the subsample \texttt{A} and \texttt{B} particles in each halo, \texttt{pos\_avg} and \texttt{vel\_avg}. The averaging procedure is more accurate for haloes containing many particles and can be used for testing the interpolation scheme as well as for placing central galaxies. The properties unique to the halo light cone catalogues are enumerated in Table~\ref{tab:halocat}.
    \item \texttt{lc\_pid\_rv.asdf} contains the particle information, i.e. the positions, \texttt{pos}, and velocities, \texttt{vel}, of the subsample \texttt{A} particles for each halo. The reason we opt for the \texttt{A} particles is that 3\% of the total output is more than sufficient for our target application of painting satellite galaxies onto the particle subsamples by adopting halo occupation techniques. The low tail of the central galaxy halo-mass distribution is above 100 particles ($\sim 2 \times 10^{11} \ M_\odot/h$) for all tracers of interest to current and near-future surveys, so the satellites are nearly always coming from halos with well above 100 particles and thus 3\% is enough to provide satellite location proxies.
\end{itemize}
The files are compressed to save space and can be loaded, read, and unpacked with the $\Abacus{CompaSO}$ reader, available at \url{https://github.com/abacusorg/abacusutils}. For more details on the compression scheme, we refer the reader to \citet{Maksimova+2021}.

\begin{table*}
  \begin{tabular}{p{6cm} p{11cm}}
  \texttt{int64\_t index\_halo} & Index of the halo into the full redshift catalogue \\
  \texttt{uint32\_t N\_interp } &  Interpolated number of particles in the halo. \\
  \texttt{float pos\_interp[3] } &  Interpolated centre of mass position of the largest L2 subhalo. \\
  \texttt{float vel\_interp[3] } &  Interpolated centre of mass velocity of the largest L2 subhalo. \\
  \texttt{float pos\_avg[3] } &  Average position of the subsample \texttt{A} and \texttt{B} particles in the halo. \\
  \texttt{float vel\_avg[3] } &  Average velocity of the subsample \texttt{A} and \texttt{B} particles in the halo. \\
  \texttt{float redshift\_interp } &  Interpolated redshift at which the light cone crosses the halo path. \\
  \texttt{int8\_t origin} & Index of the box from which the halo is taken. For haloes with merger tree information, ``0'' corresponds to original box; ``1'' or ``2'' -- copy of the box; for haloes without, ``3'' corresponds to original box; ``4'' or ``5'' -- copy of the box. 
  \end{tabular}
  \caption{Partial list of the halo field names for a single output redshift of the \Abacus{AbacusSummit} halo light cone catalogue. For a full list of the missing fields, see \citet{Maksimova+2021}. The field names shown here are reported only in the halo light cone catalogs. Note that all quantities are computed using the particles in the L2 largest subhalo of each halo.}
  \label{tab:halocat}
\end{table*}

\section{Tests and visualizations}
\label{ssec:tests}
In this section, we present tests of the validity of the halo light cone catalogues constructed using the algorithm described in the previous section. We perform a number of visual tests, consistency checks and compare theoretical predictions of the clustering with those inferred from galaxy mock light cone catalogues. Such tests were crucial, as they helped us identify implementation errors in the algorithm and also check that the interpolated positions and velocities did not introduce any unwanted features into the clustering measurements.

\subsection{Full-sky map}
As a useful visual examination of the halo light cone catalogues, we study their properties through a map of the overdensity and the gravitational tidal forces \citep{1970Ap......6..320D,2007MNRAS.375..489H,2009MNRAS.396.1815F}. The tidal tensor, defined as the Hessian of the gravitational potential, is symmetric and therefore can always be diagonalized at any point in space. In this section, we also display the eigenvalues of the tidal tensor, which inform us of the strength of the tidal forces in independent orthogonal directions.

Below we outline the method for computing the projected halo overdensity and tidal tensor. For more details on the relationship between the two-dimensional quantities computed here and the standard three-dimensional definition \cite[see][]{2016MNRAS.460..256A}. Throughout the analysis, we use the HEALPix pixelization scheme \citep{2005ApJ...622..759G} with a resolution parameter $N_{\rm side}=128$, corresponding to pixels with an area of $\sim0.21\,{\rm deg}^2$.

\begin{enumerate}
      \item We compute the overdensity field on the full sky by counting the number of haloes in each pixel $N_p$ and dividing by the average number of haloes per pixel $\bar{N}$. The field in pixel $p$ is then given by:
        \begin{equation}
          \delta_p = \frac{N_p}{\bar{N}} - 1.
        \end{equation}
      \item In order to suppress the numerical noise in the subsequent computation of derivatives on the sky, we first smooth the overdensity field using a Gaussian smoothing kernel, with standard deviation $\Delta \theta = 1^{\circ}$.
      \item We then compute the two-dimensional gravitational potential $\phi$, from the smoothed density field $\delta$ by solving Poisson's equation on the sphere, which in harmonic space states:
        \begin{equation}
          \phi_{\ell m} = -\frac{\delta_{\ell m}}{\ell(\ell+1)}.
        \end{equation}
      \item The two-dimension tidal tensor is obtained by differentiating the potential, i.e. $t_{ab}\equiv H_{ab}\,\phi$, where the Hessian operator is given by
      \begin{equation}\label{eq:cov_hess}
                \hat{H}\equiv\left(
                \begin{array}{ccc}
                 \partial_\theta^2 & \partial_\theta(\partial_\varphi/\sin\theta) \\
                 \partial_\theta(\partial_\varphi/\sin\theta) &
                 \partial_\varphi^2/\sin^2\theta+\cot\theta\partial_\theta
                \end{array}
                \right).
      \end{equation} 
      Here we compute the Hessian using the routines provided by HEALPix, which perform the derivatives in harmonic space. The tidal tensor in each pixel is then diagonalized to obtain the two eigenvalues ($\lambda_1 \geq \lambda_2$).
\end{enumerate}

\begin{figure}
    \centering
    \includegraphics[width=0.48\textwidth]{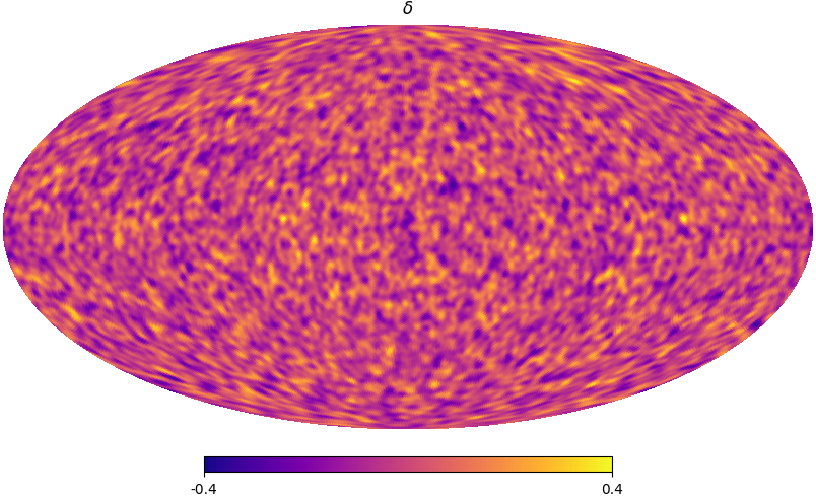}
    \includegraphics[width=0.48\textwidth]{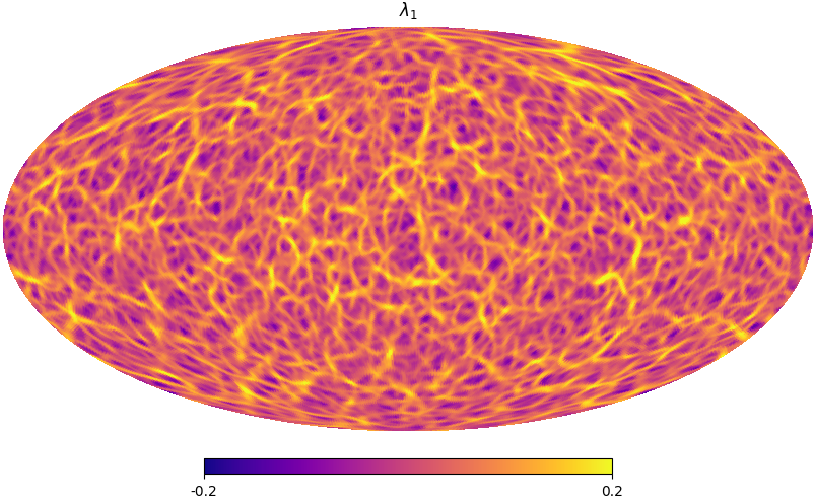}
    \includegraphics[width=0.48\textwidth]{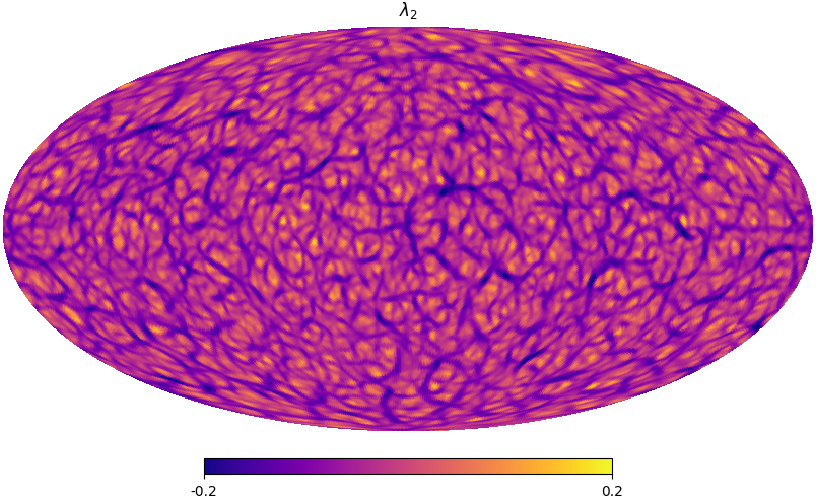}
    \caption{Density field (top panel) and eigenvalues of the two-dimensional tidal tensor of the halo light cone catalogue at $0.8 < z < 1.1$, smoothed with a 1$^\circ$ Gaussian kernel,  for the \texttt{AbacusSummit\_huge\_c000\_ph201} box with the observer placed at (0, 0, 0).}
    \label{fig:eigen}
\end{figure}

Since the observer is placed at the centre of the \texttt{huge} box, at (0, 0, 0), full-sky light cones are available for the \Abacus{AbacusSummit} \texttt{huge} simulation (until $z = 2.18$). For this exercise, we choose \texttt{AbacusSummit\_huge\_c000\_ph201}, which has $N = 8640^3$ particles and a size of $L_{\rm box} = 7500 \ {\rm Mpc}/h$. In Fig.~\ref{fig:eigen}, we show the projected halo overdensity, and the largest and smallest eigenvalues of the tidal tensor field (top and bottom, respectively). The sum of the eigenvalues maps recovers the projected overdensity. We have selected all haloes between redshifts $z = 0.8$ and $z = 1.1$, corresponding to comoving distances of 1937 and 2455 Mpc$/h$.

\subsection{Area coverage}
While the light cones of the \Abacus{AbacusSummit} \texttt{huge} boxes cover the entire sky (41253 deg$^2$), the \Abacus{AbacusSummit} \texttt{base} boxes cover a smaller fraction, since the observer is not placed at the centre of the box, but rather at the very corner of the original box (see Fig. \ref{fig:geo}). For closer redshifts ($z \leq 0.8$), the light cone information that reaches the observer comes from the original box and thus covers roughly an octant of the sky ($\sim$5300 deg$^2$), while at more distant redshifts, there are contributions from the two copies of the original box. Geometrically, the most distant available light cone shell is located at $\chi = 3990$ Mpc$/h$, corresponding to $z \approx 2.45$, for which the observer sees two patches (coming from the two copies), each of area $\sim$900 deg$^2$. Thus, at higher redshifts ($0.8 \leq z \leq 2.45$), the area coverage decreases from an octant of the sky to about 1800 deg$^2$ at $z \approx 2.45$. 

\begin{figure}
    \centering
    \includegraphics[width=0.48\textwidth]{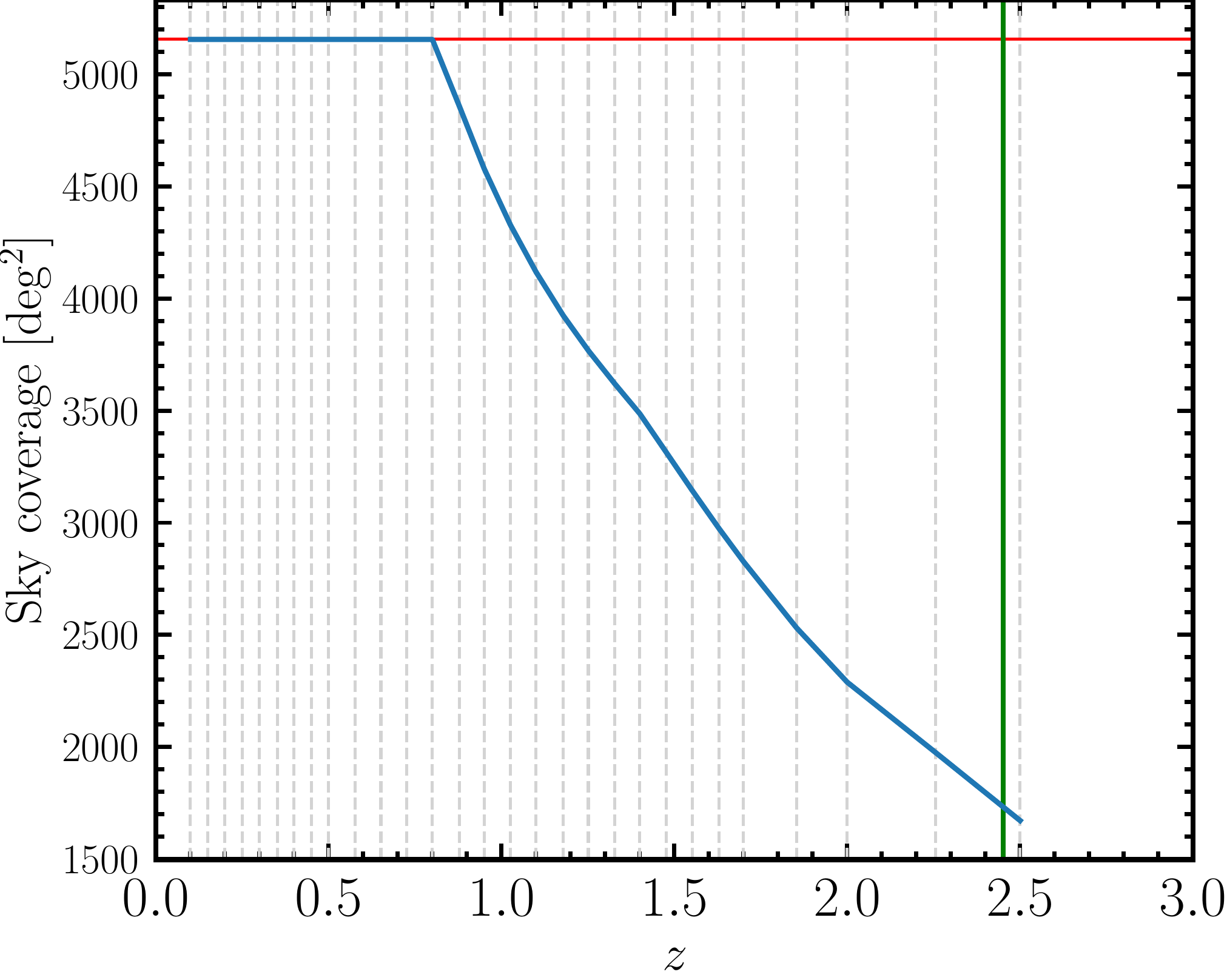}
    \caption{Sky coverage (in deg$^2$) as a function of redshift, specific to the \texttt{base} geometry. The observer is located at (-990, -990, -990), and looking at three identical copies of the box centered at (0,0,0), (0, 2000, 0) and (0, 0, 2000). All units are reported in Mpc$/h$. The red solid horizontal line corresponds to an octant of the sky, while the gray dashed vertical lines correspond to the secondary and primary redshift catalogues, at which the halo light cone catalogues are available. The green solid vertical line indicates the redshift of the furthest light cone shell, at $z \approx 2.45$, covering a total of 1800 deg$^2$ of the sky.}
    \label{fig:coverage}
\end{figure}

We illustrate this in Fig. \ref{fig:coverage}. The gray dashed vertical lines indicate the primary and secondary redshifts, at which the halo light cone catalogues are available. The green solid vertical line shows the redshift of the furthest light cone shell, at $z \approx 2.45$, while the red solid horizontal line marks an octant of the sky for reference. We note there are primary and secondary redshifts available past $z = 2.5$, but those are not present in the halo light cone catalogues, so we have not included them in the figure. The area at a distance $\chi$ from the observer (or equivalently $z$) is computed by transforming the HEALPix pixels on the sky into Cartesian coordinates located on the surface of a sphere of radius $\chi$ and calculating the number of Cartesian points that fall within the volume defined by the three copies of the box. Knowing the number of points that are located within the simulation volume, we can infer the area covered since the HEALPix pixels are equi-areal.

\subsection{Border continuity}
As a consistency check, we study the number of haloes and particles as a function of distance to the observer or equivalently as a function of redshift, $z$, which would allow us to diagnose any issues related to missing haloes or particles as well as discontinuities along the redshift borders. 

\begin{figure}
    \centering
    \includegraphics[width=0.48\textwidth]{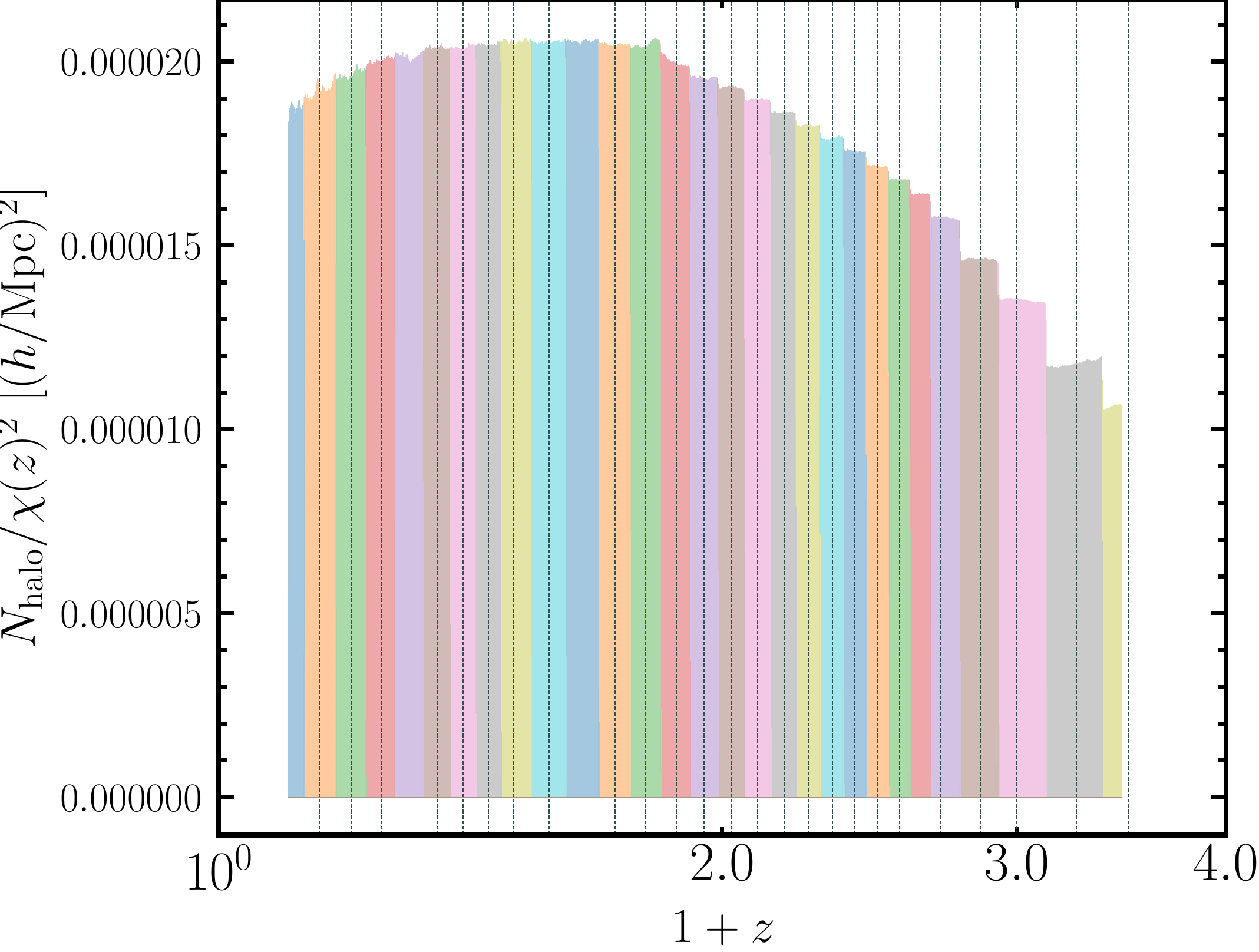}
    \caption{Number of haloes, normalized by the square of the comoving distance to the observer and the sky coverage (in rad$^2$, see Fig.~\ref{fig:coverage}), as a function of redshift. The gray dashed vertical lines correspond to the secondary and primary redshift catalogues, at which the halo light cone catalogues are available. The haloes selected for this figure contain at least 100 particles, which corresponds to a mass threshold of $M_{\rm halo} = 2.1 \times 10^{11} \ M_\odot/h$ at the \texttt{base} resolution. We expect that for higher-mass haloes, the interpolation scheme we have adopted will be more accurate, since their merger tree history is easier to track. Coincidentally, the regime above \texttt{N} $= 100$ is relevant for creating galaxy mock catalogues with different tracers.}
    \label{fig:boundaries}
\end{figure}

In  Fig.~\ref{fig:boundaries}, we show the number of haloes as a function of redshift. Since the number density of haloes in each redshift catalogue is expected to be similar, we normalize the histogram by dividing each bin by its comoving distance to the observer squared. The shape of the resulting curve roughly follows the curve of the sky coverage as a function of redshift, as expected (see Fig.~\ref{fig:coverage}). The contributions from the haloes in each redshift catalogue are painted in alternating colors to make the distinction between the catalogue boundaries clearer. In addition, we only include haloes containing at least 100 particles ($M_{\rm halo} = 2.1 \times 10^{11} \ M_\odot/h$), as this is the mass regime relevant for constructing galaxy mock catalogues for the various tracers targeted by modern galaxy surveys, e.g. emission-line galaxies (ELGs), luminous red galaxies (LRGs), quasistellar objects (QSOs), and also the interpolation mechanism adopted in our treatment is more accurate for higher-mass haloes (since lower-mass haloes have less reliable merger tree information and noisier statistics due to the lower number of particles). We have averaging over three of the \texttt{c000} simulations in order to diminish the larger variance in the number of haloes at small redshifts due to large-scale structure. 

\subsection{Accuracy of the interpolation}

Another consistency check that we perform is the measurement of the dispersion velocity of haloes, $v_{\rm disp, interp}$, belonging to the light cone catalogues as a function of their distance from the redshift boundary. In this way, we can check whether the interpolated velocities of the haloes, $\mathbf{v}_{\rm interp}$, are biased (hotter or colder) relative to the velocities of the light cone particles, which is very important when creating realistic mock catalogues. Similarly, we also check that the radial profiles of the haloes are not biased by computing the analogous quantity for the interpolated halo positions and particles, $x_{\rm disp, interp}$. The two properties are calculated for each halo as follows:
\begin{equation}
    q_{\rm disp, interp} = \sqrt{\nsum[1.6]_{i, j}\frac{(q_i^j-q_{\rm interp}^j)^2}{3 N}},
\end{equation}
where $q$ can either stand for the position $x$ or velocity $v$, $i$ is an index between 0 and $N-1$, with $N$ the total number of halo particles.

\begin{figure}
    \centering
    \includegraphics[width=0.48\textwidth]{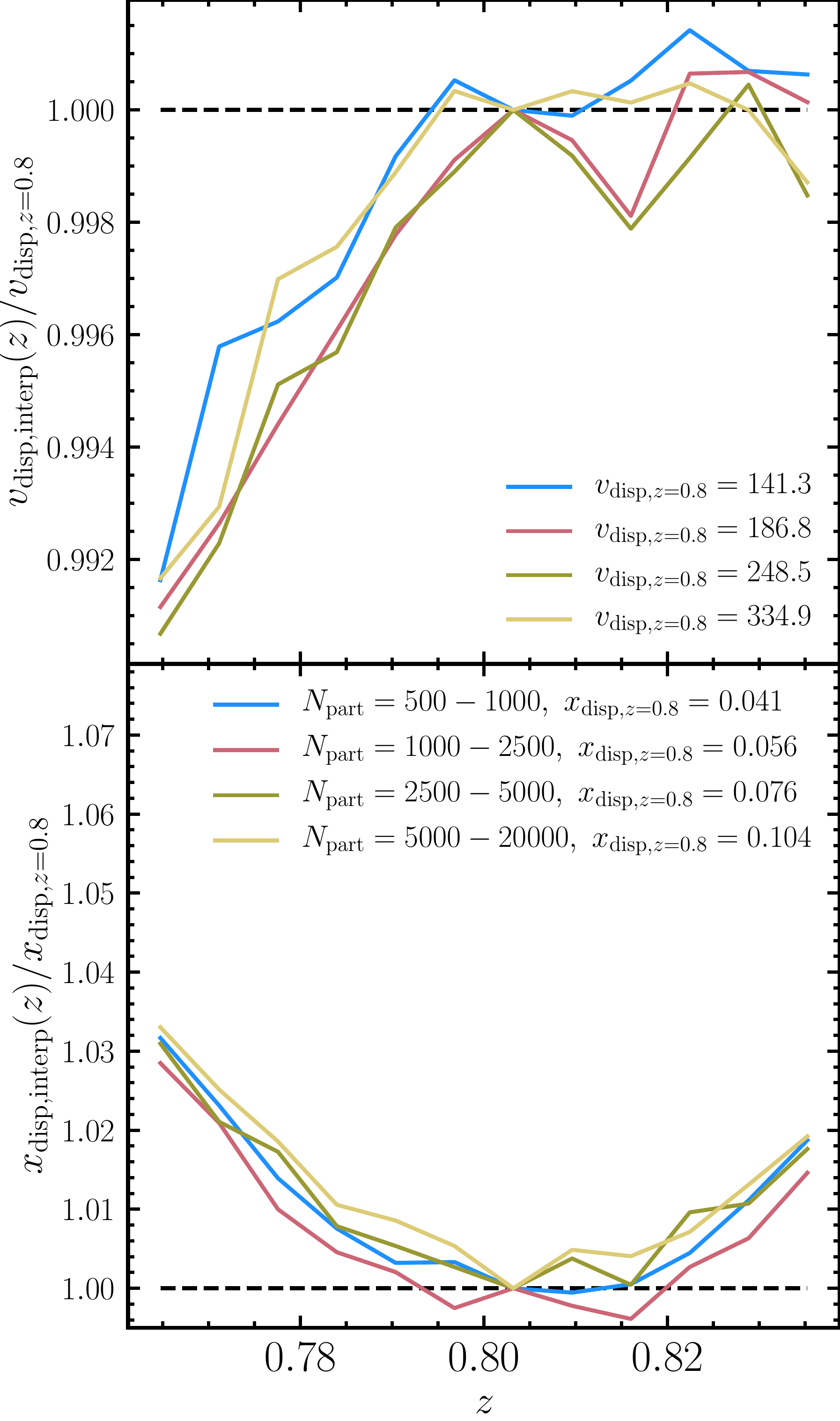}
    \caption{Dispersion velocity (top panel) and second moment (bottom panel) of haloes of different masses belonging to the light cone catalogue at $z = 0.8$ as a function of redshift. The black dashed line indicates roughly the average value of the quantity at $z = 0.8$. The shape of the $v_{\rm disp, interp}$ curve is the result of two effects. The first is the evolution of the dispersion velocity with redshift, $v_{\rm disp} \propto (1+z)^{0.35}$ \citep{2014MNRAS.440..610P}, which leads to an approximately linear increase of the curve with redshift. The second is related to the fact that the determination of which particles belong to the halo happens at $z = 0.8$. As a result, the dispersion velocities will be systemically lower, as we move away from $z = 0.8$ (in either direction). This effect also applies in the case of $x_{\rm disp, interp}$ and appears as a quadratic contribution (see bottom panel). Note that we convert all comoving coordinates to proper ones when computing $x_{\rm disp, interp}$, since haloes are expected to maintain  their proper size. These effects are small (around 2-4\%) and can be largely absorbed through the inclusion of extended HOD parameters, e.g. \texttt{s\_v}, \texttt{s}, and \texttt{s\_p} (see \ref{ssec:hod}).}
    \label{fig:interpolation}
\end{figure}

In Fig.~\ref{fig:interpolation}, we show the dispersion velocity and the analogous spatial quantity as a function of redshift. Those are computed by averaging the quantity $q_{\rm disp, interp}$ in bins of interpolated redshift (i.e. redshift at which the halo is estimated to have crossed the light cone) for haloes containing different numbers of particles. There are two effects contributing to the shape of each of these quantities. For $v_{\rm disp, interp}$, those effects come from the evolution of the dispersion velocity with redshift, empirically found to follow the relation $v_{\rm disp} \propto (1+z)^{0.35}$ \citep{2014MNRAS.440..610P}, and from the fact that the particles marked as belonging to the halo are selected at a single discrete time, the output snapshot for that redshift (i.e. at $z = 0.8$). Those particles are thus not guaranteed to be within the virial radius of the halo at $z \neq 0.8$. This leads to a systemic ``cooling'' of the particle velocities, as we move away from $z = 0.8$ and similarly, a positive quadratic increase of the second moment, $x_{\rm disp, interp}$. Furthermore, when computing the second moment, we have converted the comoving coordinates of the particles and the haloes into proper ones to account for the fact that haloes are expected to maintain their proper sizes over short periods of time. As can be seen from the figures all of these effects are small (at most around 2-4\%) and can be absorbed by enabling extra HOD parameters that model the satellite positions and velocities. The \Abacus{AbacusHOD} prescription provides three parameters to that end, \texttt{s\_v}, \texttt{s}, and \texttt{s\_p} (see \ref{ssec:hod}).


As described in Section \ref{sssec:save}, we also record the average particle positions and velocities for each halo belonging to the halo light cone catalogues, where the particles belong to subsample \texttt{A} and \texttt{B}, and their positions and velocities are taken from the light cone outputs. For haloes with a smaller number of particles ($\texttt{N} < 100$), the 10\% subsamples (i.e. coming from both \texttt{A} and \texttt{B}) might give noisy estimates of these quantities. However, for larger haloes, these two fields give a good representation of the halo bulk velocity and position. Our recommended use of the halo light cone catalogues is thus in the $\texttt{N} \gtrsim 100$ regime. Note that some haloes may have complex substructure, in which case the averaged position might be biased relative to the location of the halo nucleus, i.e. the densest particle which initiates the \Abacus{CompaSO} halo finding \citep[see][]{Hadzhiyska+2021}. In addition, we evaluate the difference between the interpolated and the averaged halo positions and velocities, $\mathbf{q}_{\rm avg}$ and $\mathbf{q}_{\rm interp}$. We find that the difference $|\mathbf{q}_{\rm avg} - \mathbf{q}_{\rm interp}|$ for haloes in the mass range of 2500 to 5000 particles is $\sim$30 kpc$/h$ for the positions and $\sim$40 km/s for the velocities. These uncertainties can be absorbed through the inclusion of additional HOD parameters that model the velocity and spatial bias of the central galaxy. For example, the \Abacus{AbacusHOD} code considers the parameter \texttt{alpha\_c}, which allow for a velocity displacement of the central galaxy relative to the peculiar velocity of the halo centre.

\subsection{Halo mass function}

{The halo mass function is an essential tool in cosmology, as dark matter haloes play a key role in modeling galaxies and galaxy clusters and performing large-scale-structure analysis. It allows us to understand the statistics of primordial matter inhomogeneities and the effects of nonlinear structure on observations through, for instance, the Sunyaev-Zeldovich effect and lensing. Another feature of the halo mass function is that it can be expressed as a universal function that relates the mass of haloes to the variance of the mass fluctuations \citep[e.g.,][]{1974ApJ...187..425P,1999MNRAS.308..119S,2001MNRAS.321..372J,2001ApJ...550L.129W,2005Natur.435..629S,2006ApJ...646..881W}.}

{As an additional test of the halo light cone catalogues, we compute the halo mass function at redshift $z = 0.8$ using both the full-box and the light cone halo catalogues:
\begin{equation}
dN_{\rm halo}(M, z) \equiv N_{\rm halo}(M, z) \, d \log M , 
\end{equation}
where we have defined $dN_{\rm halo}$ as the number of haloes in the mass range $d \log M$ between $\log M = \log{2\times 10^{11}}$ and $\log M = \log{2\times 10^{14}}$, in units of $M_\odot/h$.}

\begin{figure}
    \centering
    \includegraphics[width=0.48\textwidth]{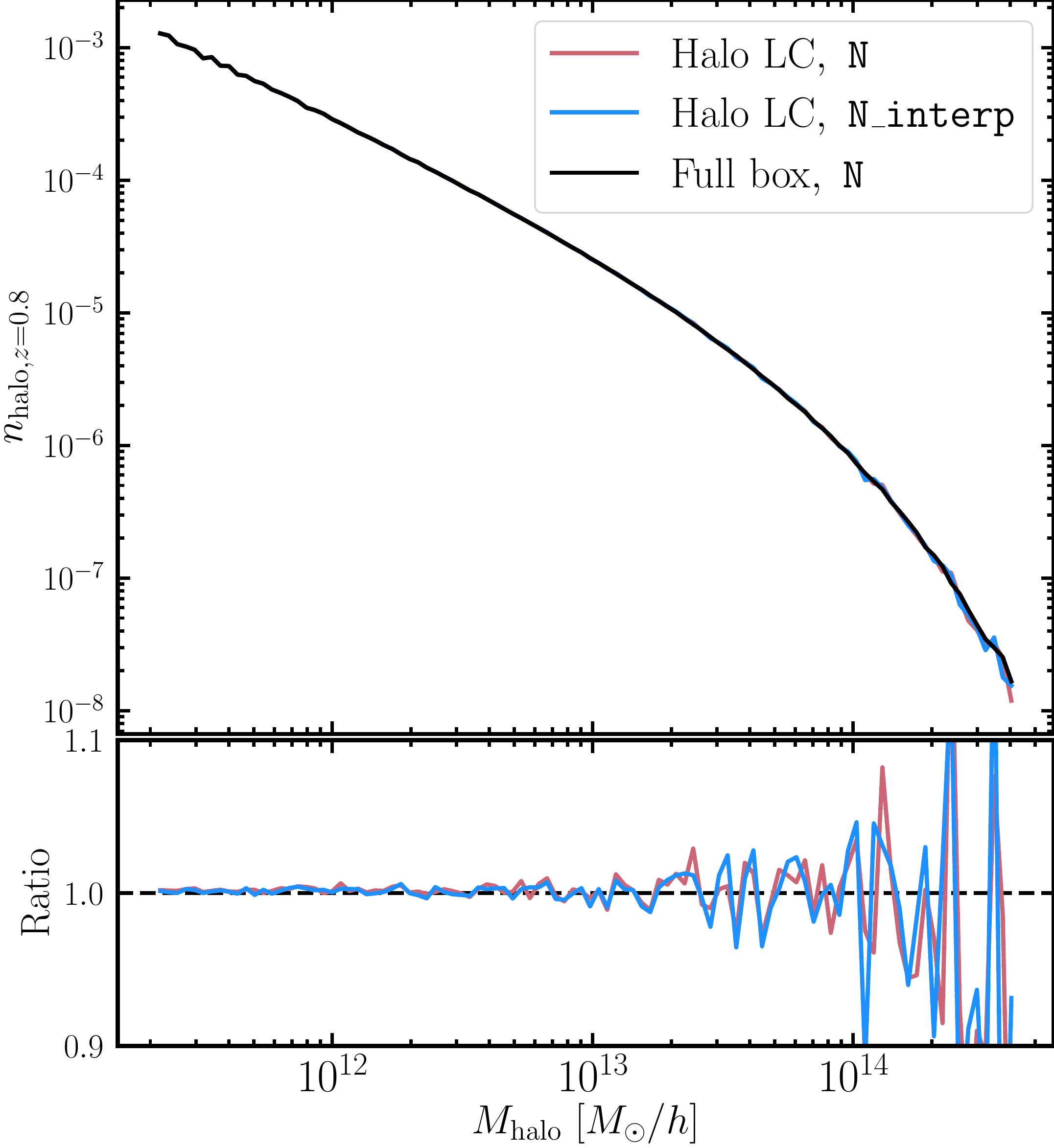}
    \caption{{Halo mass function (halo number density as a function of mass) 
    at $z = 0.8$. The black curve corresponds to the halo mass function in the full simulation box, whereas blue and red correspond to the halo mass function in the halo light cone catalogues computed using the \texttt{N\_interp} and \texttt{N} fields, respectively. The lower panel shows the ratio between the halo light cone curves and the full box curve. Halo mass is computed as ${\{\texttt{N}, \texttt{N\_{interp}}\} \times 2.1 \ 10^9} \ M_{\odot}/h$. The agreement between the halo mass functions is excellent for smaller haloes and dominated by noise for larger haloes, of which we have very few examples in the thin light cone strips at $z = 0.8$.}}
    \label{fig:hmf}
\end{figure}

{The halo mass function from the full box and the light cones is presented in Fig. \ref{fig:hmf} for the base simulation \texttt{AbacusSummit\_base\_c000\_ph006}. The \textsc{CompaSO} halo masses are defined as the total number of particles in a halo (\{\texttt{N}, \texttt{N\_{interp}}\}, see Table \ref{tab:halocat}) multiplied by the particle mass, $M_{\rm part} = 2.1 \times 10^9 \ M_{\odot}/h$. The number of particles \{\texttt{N}, \texttt{N\_{interp}}\} is taken from the cleaned \textsc{CompaSO} catalogues. We use both the non-interpolated (\texttt{N}) and interpolated (\texttt{N\_{interp}}) halo numbers of particles, shown in red and blue, respectively, as a test of the effect of interpolation on the one-dimensional halo statistics.}
{The lower segment of the panel shows the ratio of the two curves to the black full-box curve. We note that we have imposed a mass cut of $M_{\rm min} = N_{\rm min} \times M_{\rm part} = 2.1 \ 10^{11} \ M_\odot/h$, where $N_{\rm min} = 100$, since this is the regime for which the halo light cone statistics are most reliable. All three curves appear in excellent agreement with each other ($\sim$0.1\%) until $10^{13} M_\odot/h$. At higher halo masses, the ratio is dominated by noise, as the halo light cone catalogue, which has a significantly smaller volume, has fewer examples of massive haloes.}

\section{Applications}
\label{ssec:mock}
The main application of the halo light cones is in creating galaxy mock catalogues and weak lensing maps, which play a central role in developing the pipeline for current and future galaxy surveys such as DESI and \textit{Euclid}. One of the most widely used methods to conduct this so-called forward modeling of galaxy surveys is by ``painting'' them onto a halo catalogue assuming a simple empirical relation between halo mass and galaxy occupation. Since the halo occupation distribution (HOD) framework provides one of the most efficient ways of populating cosmological $N$-body simulations with galaxies and producing the many realizations required for, e.g., estimating covariance matrices \cite[e.g.][]{2009MNRAS.396...19N,2013MNRAS.428.1036M}, we have augmented the halo light cone scripts, which can be found in \url{https://github.com/abacusorg/abacusutils/tree/master/abacusnbody/hod}, with a modified HOD prescription \cite[see][and Section \ref{ssec:hod}]{Yuan+2021}.

\subsection{Emission-line galaxy catalogue}
\label{ssec:elg}
Many of the current and future cosmological surveys will target star-forming emission-line galaxies (ELGs) whose spectrum is characterized by prominent [O II] and [O III] emission lines. These galaxies will be selected by
a combination of color-color and magnitude
cuts. To model their halo occupation distribution correctly, which is vital to creating accurate mock galaxy catalogues, one can study ELG-like samples through hydrodynamical simulations. \citet{2021MNRAS.502.3599H} study the HOD of ELGs in the state-of-the-art hydrodynamical simulation IllustrisTNG \citep{2018MNRAS.475..676S,2019MNRAS.tmp.2010N,2019MNRAS.tmp.2024P} at three different time epochs: at $z = 0.8$, 1.1 and 1.4. However, in order to populate the halo light cones with ELGs that exhibit similar occupation statistics to those in IllustrisTNG, we need to parametrize these HOD curves. We do so by adopting the High Mass Quenched (HMQ) model proposed in \citet{2020MNRAS.497..581A} for the central probability of ELGs:
\begin{align}
    \left< N_{\rm cen}\right>(M_{\rm halo}) &=  2 A \phi(M_{\rm halo}) \Phi(\gamma M_{\rm halo})  + & \nonumber \\  
    \frac{1}{2Q} & \left[1+{\rm erf}\left(\frac{\log_{10}{M_{\rm halo}}-\log_{10}{M_c}}{0.01}\right) \right],  \label{eq:NHMQ}\\
\phi(x) &=\mathcal{N}(\log_{10}{ M_c},\sigma_M), \label{eq:NHMQ-phi}\\
\Phi(x) &= \int_{-\infty}^x \phi(t) \, dt = \frac{1}{2} \left[ 1+{\rm erf} \left(\frac{x}{\sqrt{2}} \right) \right], \label{eq:NHMQ-Phi}\\
A &=\frac{p_{\rm max}  -1/Q }{\max(2\phi(x)\Phi(\gamma x))}.
\label{eq:NHMQ-A}
\end{align}
The occupation statistics of the satellites are assumed to obey the standard functional form:
\begin{equation}
    \left< N_{\rm sat} \right>(M_{\rm halo}) = \left( \frac{M_{\rm halo} - \kappa M_c}{M_1}\right)^\alpha .
\end{equation}
For more details on the HMQ model and an interpretation of the various parameters, see \citet{2020MNRAS.497..581A}.

The HOD parameters that fit the IllustrisTNG ELG samples at the three redshifts $z = \{0.8, 1.1, 1.4\}$ are given below:
\begin{eqnarray}
\label{eq:params}
    p_{\rm max} = \{ 0.18, 0.13, 0.1 \}, \ Q = \{100, 100, 100\}, \\ \log M_{\rm cut} = \{11.8, 12, 12.2\}, \ \kappa = \{1.8, 1.65, 1.5\}, \nonumber \\ \nonumber   \sigma = \{0.58, 0.58, 0.58\}, \ \log M_1 = \{13.73, 13.73, 13.73\}, \\ \nonumber \alpha = \{0.7, 0.7, 0.7\}, \ \gamma = \{6.12, 5.52, 5.22\} \nonumber .
\end{eqnarray}
We have obtained these by roughly matching the HODs shown in \citet{2021MNRAS.502.3599H}. To obtain the HOD shape at intermediate redshifts and beyond $z = 1.4$, we linearly interpolate the parameters in Eq.~\ref{eq:params} at each of the available halo light cone redshifts. For $0.45 < z < 0.8$, we assume that the HOD shape is identical to that at $z = 0.8$.

\subsubsection{Redshift space distortions}
An advantage of having mock catalogues on the light cone is that they can be used to construct forward models of various large-scale structure tracers, which are invaluable for modern galaxy redshift surveys such as DESI and \textit{Euclid}. Measuring redshift space distortions (RSD) has become a standard application of galaxy surveys \cite[e.g.][]{2011MNRAS.415.2876B,2015MNRAS.449..848H,2017MNRAS.470.2617A,2017A&A...604A..33P}. Studies of RSD clustering statistics exist in both configuration space and Fourier space, each coming with its own benefits and challenges. Here, we will concentrate on the real-space statistics.

The clustering anisotropy introduced due to galaxy velocities may be described by a multipole expansion of the correlation function with respect to the local line of sight \citep{1994MNRAS.267..785C,1998ASSL..231..185H}, which permits a powerful compression of the information. To estimate the multipoles, one needs to measure the correlation as a function of separation $r$ and angle $\mu$. For a catalogue with an arbitrary redshift distribution and arbitrary survey boundaries, in addition to counting pairs of the galaxy tracers, one also needs to consider pair counts of randomly distributed objects, exhibiting the same redshift distribution and survey boundaries. The Landy-Szalay (LS) estimator \citep{1993ApJ...412...64L} combines all possible correlations between data, $D$, and randoms, $R$, to calculate the underlying correlation function between two arbitrary tracers in a nearly optimal way:
\begin{equation}
\xi^s(r,\mu) = \frac{\langle D_1 D_2 \rangle - \langle D_1 R_2 \rangle - \langle D_2 R_1 \rangle + \langle R_1 R_2 \rangle}{\langle R_1 R_2 \rangle}, 
\label{eq:ls}
\end{equation}
where the angle brackets denote normalized pair counts at separation $r$ and angle $\mu$. We can decompose the redshift-space correlation function into multipoles using the Legendre polynomials $P_\ell(\mu)$ as:
\begin{equation}
 \xi_\ell(r) = \int_0^1\xi^s(r,\mu)(1+2\ell)P_\ell(\mu)\mathrm{d}\mu,
\label{eq:xi_l}
\end{equation}
where the correlations in Eq.~\ref{eq:ls} need to be computed in bins of $r$ and $\mu$ before taking the ratio.

\begin{figure*}
    \centering
    \includegraphics[width=0.48\textwidth]{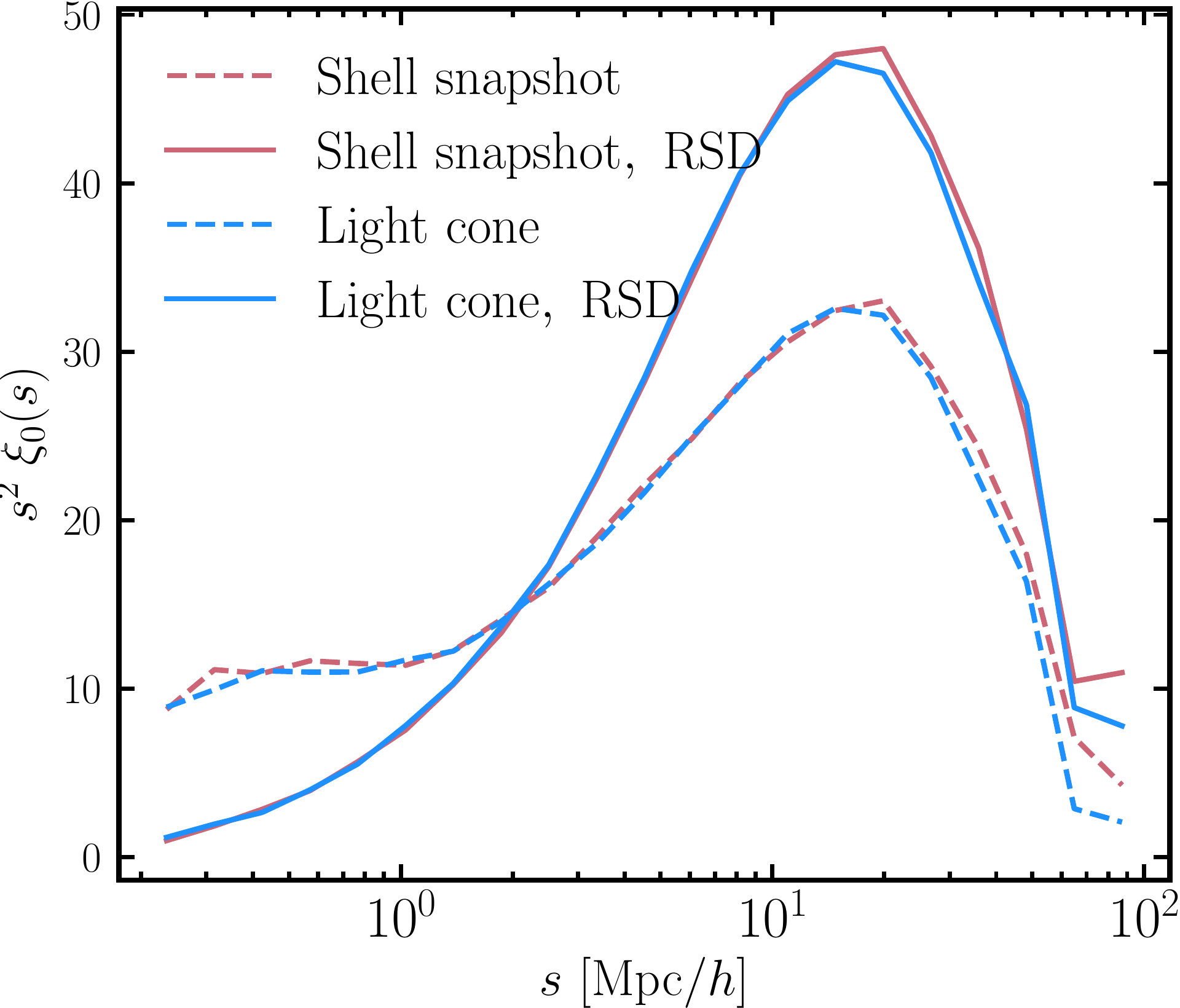}
    \includegraphics[width=0.48\textwidth]{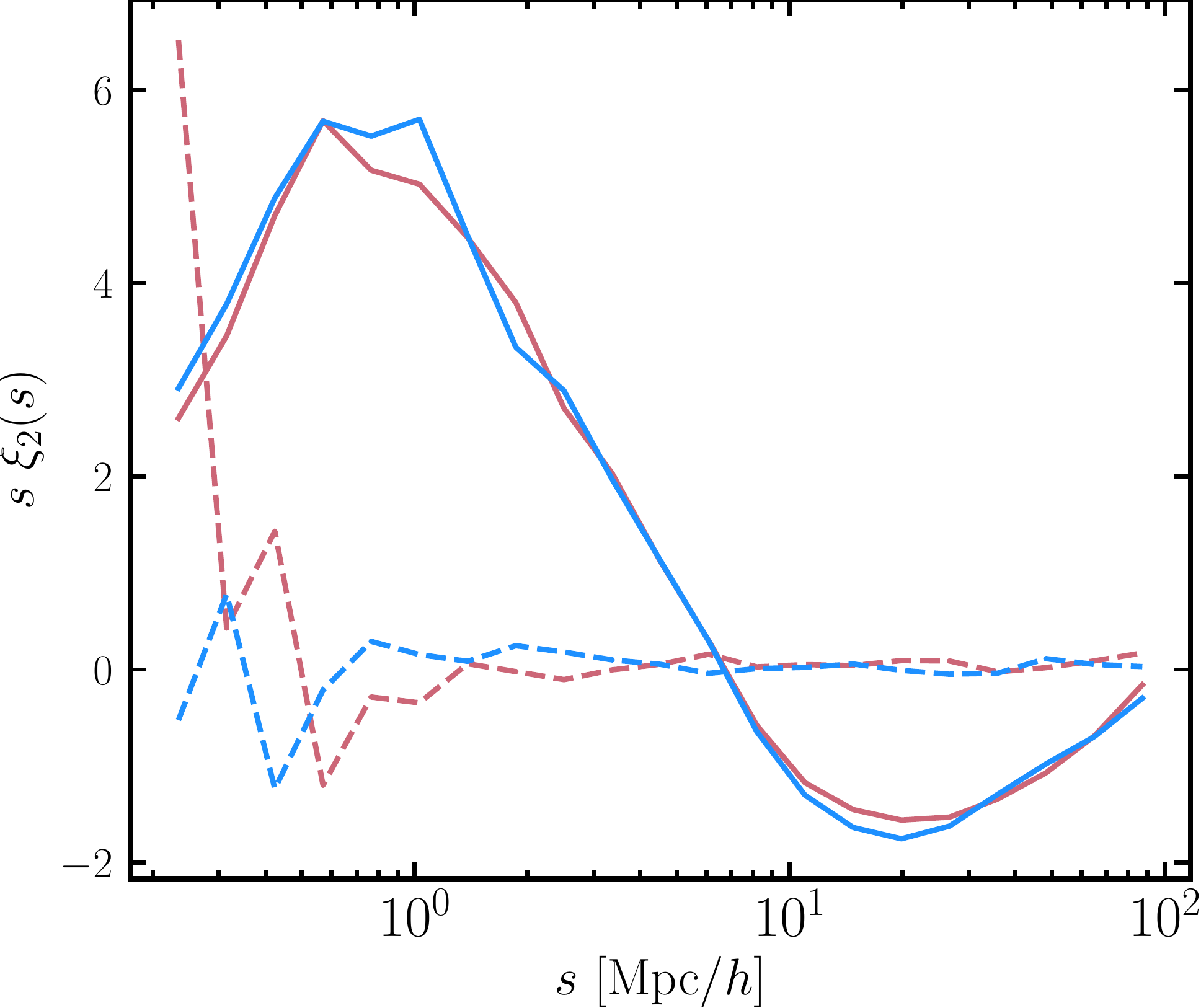}
    \caption{The monopole (left panel) and quadrupole (right panel) of the ELG correlation function at $z = 0.5$ with and without RSD effects (solid and dashed lines, respectively). In blue, we show the prediction from the halo light cone catalogues, whereas in red, we show the result from the full snapshot catalogues. The full-snapshot sample is constructed by selecting a spherical shell of thickness and radius equal to that of the light cone catalogue at that redshift ($\sim 140 \ {\rm Mpc}/h$) and applying RSD effects in the direction of the observer at (-990, -990, -990) ${\rm Mpc}/h$. The agreement between the halo light cone and the snapshot shell for both the monopole and the quadrupole is very good, which serves as an implicit validation of the halo light cone catalogues. With the RSD effects are switched off, the monopole $\xi_0$ becomes the real-space correlation function $\xi(r)$, so the quadrupole vanishes. This is confirmed in the right panel, where we see that the quadrupole signal is consistent with zero for both samples.}
    \label{fig:xi_l}
\end{figure*}

To test the mock catalogues obtained on the light cone, we compute the monopole and quadrupole, following the prescription above. In addition, we construct a second sample from the ``stationary'' snapshot catalogues by selecting the galaxies within a spherical shell of average radius $\sim 1330 \ {\rm Mpc}/h$ and thickness $\sim 140 \ {\rm Mpc}/h$, centered at the observer's location, (-990, -990, -990) ${\rm Mpc}/h$. The shell parameters are chosen such that they correspond to the thickness and radius of the halo light cone catalogue at that redshift. RSD effects have been applied to the galaxies prior to the shell selection. We also study the effect on the galaxy correlation function when the RSD effects are switched off. Fig.~\ref{fig:xi_l} demonstrates that the agreement between the shell snapshot catalogue and the light cone catalogue is very good on all scales for both the monopole and the quadrupole. In addition, we find that the quadrupole signal vanishes when the RSD effects are neglected due to the isotropical distribution of the galaxies.

\subsection{Convergence maps}
\label{ssec:cmb}
In this section, we describe our method for computing the convergence maps from the particle light cones. 
Adopting the Born approximation and assuming flat space, the convergence field for lensing distortions is given by
\begin{equation}
\kappa(\theta) = {3 H_0^2\Omega_m\over{2c^2}}~\int ~d \chi~\delta(\chi,\theta)
{(\chi_s-r)\chi\over{\chi_s~a}} 
\label{eq:kappa}
\end{equation}
where $H_0 = 100 h$ km/s/Mpc is the Hubble constant, $\Omega_m$ is the energy density of matter, $c$ is the speed of light,  $\delta$ is the 3-D matter overdensity at
radial distance $\chi(a)$ (for a corresponding scale factor $a$), $\theta$ is the angular position, and $\chi_s$ is the
distance to the lensing source(s).

This equation can be discretized and used to compute the convergence map in an $N$-body simulation by adding up the dark-matter shells in the light cone weighted by the weak-lensing kernel at each redshift. This can be done as
follows \citep{2008MNRAS.391..435F}:
\begin{equation}
    \kappa(i) = {3H_0^2\Omega_{cb}\over{2c^2}}~\sum_j ~\delta(i,j)~{(\chi_s-\chi_j)\chi_j\over{\chi_s a_j}}~d\chi_j 
\end{equation}
where $i$ indicates the pixel position in the sky and $j$ the radial bin index (at distance $\chi_j$ of width $d\chi_j$) into which we have sliced the simulation. Note that since \Abacus{AbacusSummit} uses a basic prescription for neutrinos, modeling them as a smooth, non-clustering matter component \citep{Maksimova+2021}, we need to consider the contribution only from the gravitational components, i.e. baryons and cold dark matter, $\Omega_{cb} = \Omega_b + \Omega_{c}$. The \Abacus{AbacusSummit} treatment of neutrinos captures accurately the suppression on small scales, but does not account for the neutrino clustering on large scales. However, this is a secondary effect and does not matter for most applications relevant for galaxy surveys. Denoting the number of particles in pixel $i$ and slice $j$ as $N_{ij}$, we compute the overdensity as:
\begin{equation}
    \delta(i,j) = \frac{\rho(i,j)}{{\bar{\rho}}}-1
\end{equation}
where $\bar{\rho}= \langle \rho(i,j) \rangle$ is the mean density, which we compute analytically as $(N_{\rm part}/L_{\rm box}^3) {\Delta \Omega ~\chi_j^2~ d\chi_j}$, and
\begin{equation}
\rho(i,j) = \frac{N_{i,j}}{{dV_j}} = \frac{N_{i,j}}{{\Delta \Omega ~\chi_j^2~ d\chi_j}}
\end{equation}
where $\Delta\Omega$ is the area of each pixel. 

\subsubsection{Cross-correlation of ELGs with the CMB}
During their travel from the last scattering surface to the observer, CMB photons interact with the matter inhomogeneities and information about the large-scale clustering of the Universe gets imprinted onto the CMB temperature and polarization anisotropies. One of the ways in which matter and the CMB interact is through gravitational lensing, which causes small but coherent deflections of the path of CMB photons. Careful modeling of this effect enables the reconstruction of the gravitational potential integrated along the line of sight \citep{2002ApJ...574..566H,2003PhRvD..68h3002H}. A relatively novel approach to studying CMB lensing is through cross-correlations with other tracers of large-scale structure. This allows us to constrain the evolution of dark matter density fluctuations and dark energy at the dawn of cosmic acceleration. Cross-correlation measurements also encode information about the cosmic bias and the effective halo masses associated with the tracer populations. The advantage of gaining that information through the cross-power spectra as opposed to or in addition to the auto-power spectra is that cross-correlation measurements do not suffer from systematics that are not correlated between the two data sets. Thus, they can uncover unforeseen systematics in either dataset as well as constrain the galaxy bias in an independent fashion.

To compute the CMB convergence field in \Abacus{AbacusSummit}, we can assume that the distance to the lensing source is constant and equal to $\chi_s (z_{\rm rec}) \approx 13873$ Mpc, which corresponds to $z_{\rm rec} = 1089.3$ at the fiducial cosmology and substitute it in Eq.~\ref{eq:kappa}. The \Abacus{AbacusSummit} light cone geometry provides complete light cone information for $0.1 \leq z \leq 2.45$ covering 1800 deg$^2$ (split between two patches of equal area). All subsequent measurements of the angular power spectra discussed in the text are calculated on the masked map of area 1800 deg$^2$, corrected for the fraction of sky covered. Note that this does not span the entire active domain of the CMB lensing kernel, so we need to be careful when comparing simulation observables with theory. To this end, we perform the line-of-sight integration necessary for the theoretical predictions only within the redshift range $0.1 \leq z \leq 2.45$. 

Having obtained the CMB convergence map, we can compare it with the theoretically estimated angular power spectrum from \texttt{pyccl} \citep{1812.05995} given by:
\begin{equation}
    C_l(\kappa) = \frac{9H_0^4\Omega_m^2}{{4c^4}}~\int ~d\chi~ P(k,z) \frac{(\chi_s-\chi)^2}{{\chi_s^2~a^2}}
\end{equation}
where $P(k,z)$ is the three-dimensional density power spectrum in the simulation at redshift $z$ evaluated at $k=l/\chi$ in the Limber approximation \citep{1953ApJ...117..134L}, valid for $l>10$ within a few percent accuracy \citep[see e.g.,][]{2002PhR...367....1B}.  

In addition to the auto-power spectrum of the CMB convergence field, we can calculate the cross-correlation signal between any large-scale structure tracer and the CMB convergence field. Similarly to Section \ref{ssec:elg}, we will work with an ELG sample, as ELGs will take up a sizable portion of the objects studied by many current and future cosmological galaxy surveys such as DESI and \textit{Euclid}. To create the ELG mock galaxy catalogue on the halo light cone, we follow the same approach as in Section \ref{ssec:elg}. The redshift dependence of the HOD parameters is obtained by linearly interpolating (and extrapolating assuming constant derivatives) between the three redshifts for which we have studied ELGs using the hydrodynamical simulation IllustrisTNG \citep{2021MNRAS.502.3599H}: $z = 0.8, \ 1.1, \ {\rm and} \ 1.4$. Assuming that the observed galaxy sample has a Gaussian redshift distribution $N(z)$ centered at $z = 1.1$ and with standard deviation of $\Delta z = 0.15$, we downsample the galaxies in our mock catalogue. The galaxy bias passed to \texttt{pyccl} is assumed to be redshift-dependent and given by $b(z) = 1.14/D(z)$, where $D(z)$ is the growth factor at redshift $z$.

\begin{figure}
    \centering
    \includegraphics[width=0.48\textwidth]{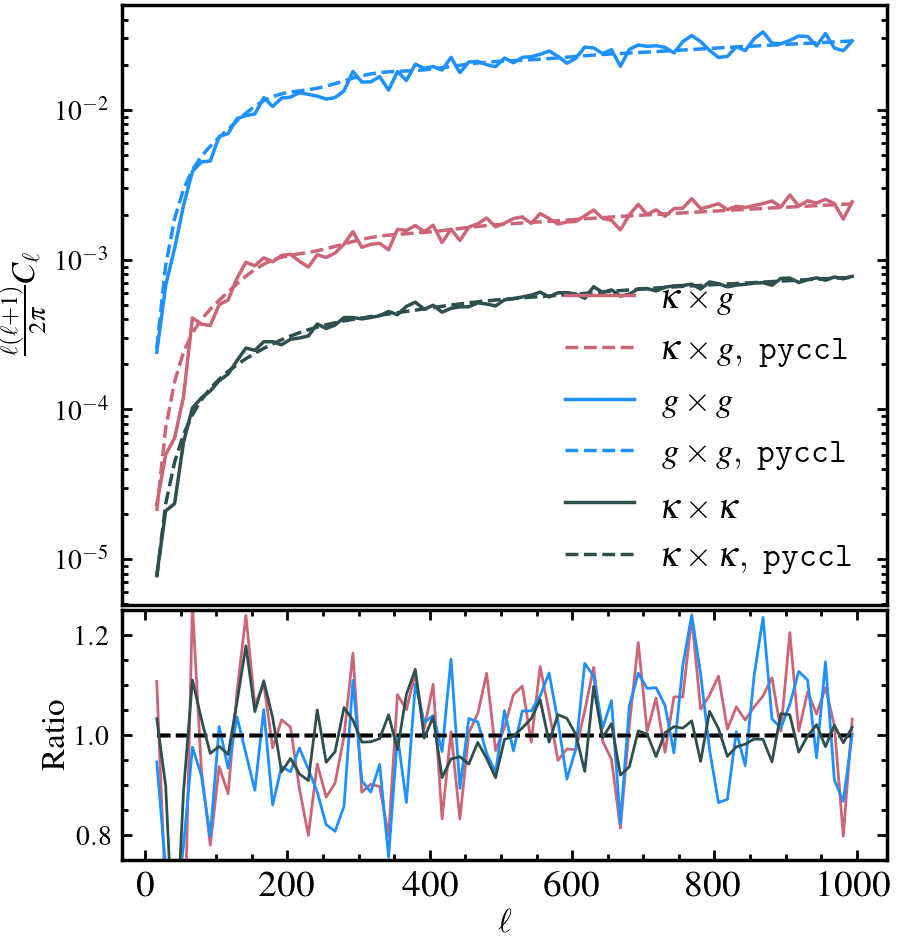}
    \caption{Auto- and cross-power spectra for the CMB convergence field and an ELG sample with $N(z) = \mathcal{N}(\mu = 1.1, \sigma=0.15)$ described in Section \ref{ssec:elg}. In red, we show the cross-power spectrum between galaxies and the CMB convergence field, while in blue and gray, we show the auto-power spectra for the galaxies and convergence field, respectively. The agreement between the theoretical predictions and the computed power spectra is very good for $\ell > 100$: $\kappa \times \kappa$ and $g \times g$ differ respectively by 0.1\% and 0.1\% from the \texttt{pyccl} curve, while $\kappa \times g$ deviates by about 1.7\%, which is within the theoretical error.}
    \label{fig:cls_cmb_elg}
\end{figure}

A comparison between the measured power spectra and their respective theoretical predictions are shown in Fig.~\ref{fig:cls_cmb_elg}. We find that the agreement between theory and observations for $\ell > 100$ is very good for all three combinations: galaxy-galaxy ($g \times g$), galaxy-convergence field ($\kappa \times g$), and convergence-convergence ($\kappa \times \kappa$). In particular, the \texttt{pyccl} prediction for $C_\ell^{\kappa \kappa}$ differs by about 0.1\% from the \Abacus{AbacusSummit} estimate., while $C_\ell^{gg}$ and $C_\ell^{\kappa g}$ are consistent with theory at the 0.1\% and 1.7\% level, respectively. This difference is within the margins of the theoretical error. We note that since light cone information is available only for $0.1 \leq z \leq 2.45$, we set the CMB lensing kernel to zero outside of that redshift range.

\begin{figure}
    \centering
    \includegraphics[width=0.48\textwidth]{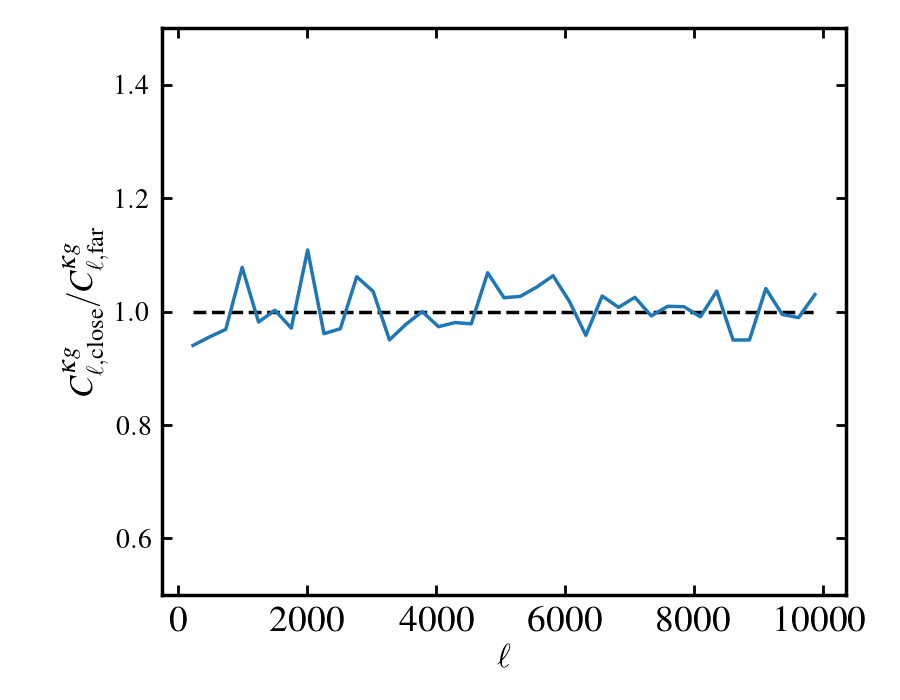}
    \caption{Ratio of the galaxy-CMB lensing cross-correlation power spectra for two ELG subsamples. The first subsample (``close'') contains galaxies with light-cone crossing redshifts, $z_\ast$, satisfying $z_{i-1/4} < z_\ast <= z_{i+1/4}$, while the second one (``far'') contains the rest of the galaxies in the ELG sample from Fig.~\ref{fig:cls_cmb_elg}. The ratio is very close to one, suggesting that the interpolation effects introduced by our algorithm (see Section~\ref{sec:light}) are negligible even at these very small scales.}
    \label{fig:cl_kg_ratio}
\end{figure}

Future studies combining early Universe probes such as CMB lensing with the late-time galaxy distribution will measure with great accuracy the very small scales around $\ell \sim 10000$. Therefore, it is important that the halo light cone catalogues do not suffer from substantial biasing of the galaxy-matter cross-correlation, resulting from the interpolation techniques we have adopted (see Section~\ref{sec:light}). The haloes whose positions and velocities are likely to be the most strongly affected by interpolation effects have light-cone crossing redshifts, $z_\ast$, near the mid-point between any two redshift catalogues. We, thus, define two galaxy subsamples using the ELG main sample used in Fig.~\ref{fig:cls_cmb_elg}. The first subsample (``close'') consists of all galaxies that have $z_{i-1/4} < z_\ast <= z_{i+1/4}$ for all redshift catalogues, $z_i$, whereas the second one (``far'') contains the rest of the galaxies in the sample. We then compute the galaxy-convergence field cross-correlation power spectrum, $C_\ell^{\kappa g}$ and show the ratio for the two subsamples in Fig.~\ref{fig:cl_kg_ratio}. We do not see significant deviations of this ratio from one, indicating that the interpolation procedure does not affect the smallest scales of the galaxy .

\section{Conclusions}
\label{sec:conc}

The amount of cosmological information we extract from current and future cosmological surveys depends crucially on the techniques we adopt to emulate realistic galaxy catalogues through simulations. These so-called mock catalogues are useful for developing analysis tools, assessing incompleteness, making predictions, and computing accurate covariance matrices.

In this paper, we have described the procedure for obtaining halo light cone catalogues from the \Abacus{AbacusSummit} simulation. We first create a halo light cone from the merger trees by calculating the interpolated positions and velocities of each halo at the redshift at which it crosses the observer’s lightcone. We then associate the particles belonging to these haloes with the particle light cone outputs. For the \Abacus{AbacusSummit} \texttt{base} simulations with a mass resolution of $2.109 \times 10^9 \ M_\odot/h$ and a box of size $L_{\rm box} = 2$ Gpc$/h$, the halo light cone catalogue covers an octant of the sky extending to $z \sim 0.8$, making use of a single copy of the box, and about 1800 deg$^2$ extending to $z \approx 2.45$ when we utilize two additional copies. For the \Abacus{AbacusSummit} \texttt{huge} simulations, the light cone comes from a single copy of the box. This provides a full-sky light cone to the the half-distance of the box (3.75 Gpc$/h$, corresponding to $z = 2.18$), and further toward the eight corners (e.g. half the sky till $z = 3.2$). The final product can then be populated with galaxies using the augmented \Abacus{AbacusHOD} prescription. We recommend using the halo light cone products in the halo mass regime of $M_{\rm halo} > 2.1 \times 10^{11} \ M_\odot/h$, corresponding to haloes containing 100 particles or more.

We perform multiple tests to check the validity of the generated halo light cone catalogues. In particular, we find that there are no evident discontinuities at the boundaries between the different redshift catalogues and that the number of haloes as a function of redshift is consistent with the expectation of nearly constant halo density, accounting for the dependence of the sky coverage on redshift. In addition, we visualize the \texttt{huge} box for $0.8 < z < 1.1$ and recover the full-sky projected density field and tidal tensor, finding no unphysical features. We further study the accuracy of the interpolation technique we have adopted and find that the difference between the averaged particle velocity or position and the interpolated one is small ($|\mathbf{x}_{\rm interp} - \mathbf{x}_{\rm avg}| = 30 \ {\rm kpc}/h$ and $|\mathbf{v}_{\rm interp} - \mathbf{v}_{\rm avg}| = 40 \ {\rm km/s}$ for haloes containing 2500 to 5000 particles at $z = 0.8$). These differences can have different causes: e.g., haloes that have rich substructure will report biased $\mathbf{x}_{\rm avg}$ and $\mathbf{v}_{\rm avg}$, whereas smaller haloes might be harder to track reliably through time in particular as they fly by large clusters. We also study the dispersion velocities and second moments of haloes and find that while there are percent-level effects coming from the fact that halo membership of the particles is reported at only a handful of redshifts. Luckily, these effects can be absorbed through extra HOD parameters.

Galaxies can be assigned to the halo light cone catalogues using a modified version of \Abacus{AbacusHOD}, detailed in \citet{Yuan+2021}, which takes as input the halo light cone catalogues rather than the full snapshot outputs. \Abacus{AbacusHOD} offers a sophisticated prescription of the standard HOD method that generalizes it with various halo-scale physics such as satellite distribution, velocity bias, closest approach distance, and assembly bias in the form of concentration and environment. We generate mock catalogues of emission-line galaxies (ELGs) on the light cone and compare our prediction for the compressed galaxy clustering statistics (i.e. the monopole and quadrupole of the two-point function) at $z = 0.5$ with the full snapshot result. We find that our mock catalogues manage to recover the observables with very good accuracy. This application is based on creating a single galaxy mock catalogue at a fixed cosmology, but in order to place tight constraints on cosmological parameters through galaxy surveys, we need to compute high-dimensional covariance matrices, which requires us to generate thousands of mock catalogues. This could be achieved by applying approximate fast methods for generating halo catalogues \cite[e.g.][]{2013MNRAS.428.1036M,2013MNRAS.433.2389M,2013JCAP...06..036T,2014MNRAS.437.2594W,2015MNRAS.450.1856A,2015MNRAS.452..686C,2015MNRAS.450.1836K}, but a significant advantage of the \Abacus{AbacusHOD} model \citep{Yuan+2021} is that it is highly  optimized and therefore well-suited for this task.

In addition, we showcase a cross-correlation study between the galaxy clustering of ELGs at $z \sim 1.1$ and CMB lensing. The convergence maps of the CMB are computed using all available light cone outputs between $0.1 < z < 2.45$, accounting for the way in which neutrinos are treated in \Abacus{AbacusSummit} (i.e. as a non-clustering, smooth component). We find that the auto-correlation of the convergence maps agrees with the theoretical prediction from \texttt{pyccl} at the 0.1\% level. As for the slightly more noisy measurements of the galaxy-CMB lensing and galaxy-galaxy auto-power spectra, we find those to be consistent with theory within 1.7\% and 0.1\%, where we have assumed a simple form of the redshift-dependent bias, i.e. $b(z) = 1.14/D(z)$.

In the near future, multiple surveys will be mapping the distribution of galaxies, so developing efficient galaxy population tools that reproduce their clustering properties with a high degree of fidelity will become crucial to advancing precision cosmology. Applying these tools to simulations, we can obtain mock catalogues and build high-precision covariance matrices for quantifying the uncertainties in estimates of cosmological parameters. Such an endeavor could potentially bridge important gaps in light of future galaxy surveys and truly enable us to make remarkable subpercent inferences about the makeup of our Universe.

\section*{Acknowledgements}
We thank Sihan Yuan and Tanveer Karim for many illuminating discussions. {We are also grateful to the anonymous referee for their helpful input and comments.}

This work has been supported by NSF AST-1313285, DOE-SC0013718, and NASA ROSES grant 12-EUCLID12-0004.
DJE is supported in part as a Simons Foundation investigator. 
LHG is supported by the centre for Computational Astrophysics at the Flatiron Institute, which is supported by the Simons Foundation.  
SB is supported by Harvard University through the ITC Fellowship.

This research used resources of the Oak Ridge Leadership Computing Facility, which is a DOE Office of Science User Facility supported under Contract DE-AC05-00OR22725. 
Computation of the merger trees used resources of the National Energy Research Scientific Computing centre (NERSC), a U.S. Department of Energy Office of Science User Facility located at Lawrence Berkeley National Laboratory, operated under Contract No. DE-AC02-05CH11231.
The \textsc{AbacusSummit} simulations have been supported by OLCF projects AST135 and AST145, the latter through the Department of Energy ALCC program.

We would like to thank the OLCF and NERSC staff for their highly responsive and expert assistance, both scientific and administrative, during the course of this project.


\section*{Data Availability}

The simulation data is available as part of \Abacus{AbacusSummit} and is subject to the academic citations described at \url{https://abacussummit.readthedocs.io/en/latest/citation.html}.

Data access is available through OLCF's Constellation portal. The persistent DOI describing the data release is \href{https://www.doi.org/10.13139/OLCF/1825069}{10.13139/OLCF/1825069}. Instructions for accessing the data are given at \href{https://abacussummit.readthedocs.io/en/latest/data-access.html}{https://abacussummit.readthedocs.io/en/latest/data-access.html}.




\bibliographystyle{mnras}
\bibliography{refs} 



\appendix



\bsp	
\label{lastpage}
\end{document}